\documentclass[12pt,onecolumn,journal]{article}
\usepackage{paralist}
\usepackage{graphics}
\usepackage{epsfig}
\usepackage{graphicx}
\usepackage[normalem]{ulem}
\usepackage{float}
\usepackage[caption=false]{subfig}
\graphicspath{ {figures/} }
\usepackage{mathtools}
\usepackage{bibentry}
\usepackage[all,cmtip]{xy}
\usepackage{a4wide}
\usepackage{amsthm, amssymb,dsfont}
\usepackage{graphics,graphicx}
\usepackage{epstopdf}
\usepackage{amsmath}
\usepackage{cleveref}
\usepackage{color}

\usepackage{algorithmicx,algorithm}
\usepackage{algpseudocode}

\usepackage{enumerate}
\usepackage{cite}
\usepackage{amsfonts,amssymb}
\usepackage{setspace}

\newtheorem{theorem}{Theorem}[section]

\newtheorem{proposition}[theorem]{Proposition}
\newtheorem{lemma}[theorem]{Lemma}
\newtheorem{remark}[theorem]{Remark}
\newtheorem{example}[theorem]{Example}
\newtheorem{corollary}[theorem]{Corollary}

\def\ed{ \end{document} }

\usepackage[noblocks]{authblk}
\author[1]{Kelvin C.K. Chan}
\author[1]{Raymond H. Chan\thanks{Research supported by HKRGC Grants No. CUHK14306316,  HKRGC CRF Grant C1007-15G, HKRGC AoE Grant AoE/M-05/12,
CUHK DAG No. 4053211, and CUHK FIS Grant No. 1907303. }}
\author[2]{Mila Nikolova\thanks{Research supported by the French Research Agency (ANR)
under grant No ANR-14-CE27-001 (MIRIAM)
and by the Isaac Newton Institute for Mathematical Sciences for
support and hospitality during the programme Variational Methods and Effective Algorithms for Imaging and Vision,
EPSRC  grant  no  EP/K032208/1.}}
\affil[1]{\footnotesize Department of Mathematics, The Chinese University of Hong Kong,
Shatin, Hong Kong, China (kelvinckchan@outlook.com and rchan@math.cuhk.edu.hk)}
\affil[2]{\footnotesize Centre de Math\'ematiques et de Leurs Applications, ENS de Cachan, 94235 Cachan Cedex, France (nikolova@cmla.ens-cachan.fr)}

\begin{document}

\title{A Convex Model for Edge-Histogram Specification with Applications to Edge-preserving Smoothing}

\date{}
\maketitle

\begin{abstract}
The goal of edge-histogram specification is to find an image whose edge image has a histogram that matches a given edge-histogram as much as possible. Mignotte has proposed a non-convex model for the problem [M. Mignotte. An energy-based model for the image edge-histogram specification problem. \textit{IEEE Transactions on Image Processing}, 21(1):379--386, 2012]. In his work, edge magnitudes of an input image are first modified by histogram specification to match the given edge-histogram. Then, a non-convex model is minimized to find an output image whose edge-histogram matches the modified edge-histogram.  The non-convexity of the model hinders the computations and the inclusion of useful constraints such as the dynamic range constraint. In this paper, instead of considering edge magnitudes, we directly consider the image gradients and propose a convex model based on them. Furthermore, we include additional constraints in our model based on different applications. The convexity of our model allows us to compute the output image efficiently using either Alternating Direction Method of Multipliers or Fast Iterative Shrinkage-Thresholding Algorithm. We consider several applications in edge-preserving smoothing including image abstraction, edge extraction, details exaggeration, and documents scan-through removal. Numerical results are given to illustrate that our method successfully produces decent results efficiently.
\end{abstract}
%%%%%%%%%%%%%%%%%%%%%%%%%%%%%%%%%%%%%%%%%%%%%%%%%%%%%%%%%%%%%%%%%%%%%%%%%%%%%%%%
%%%%%%%%%%%%%%%%%%%%%%%%%%%%%%%%%%%%%%%%%%%%%%%%%%%%%%%%%%%%%%%%%%%%%%%%%%%%%%%%
\section{Introduction}
\label{S:1}
Histogram specification is a process where the image histogram is altered such that the histogram of the output image follows a prescribed distribution. It is one of the many important tools in image processing with numerous applications such as image enhancement \cite{lee2017color,lim2015new,wang1999image}, segmentation \cite{yu2010otsu,tobias2002image,thomas2008image} among many others.

The goal of edge-histogram specification is to find an image whose edge image has a histogram that matches a given edge-histogram as much as possible. In \cite{mignotte2012energy}, Mignotte proposed a non-convex model for the problem. Given a discrete input image $I$ of size $m$-by-$n$, let $\mathbf{r}\in\mathbb{R}^k$ be a vector storing the pairwise differences
\begin{equation}
	\label{mignotte_r}
	r_{s,t} := |I_s - I_t|, \quad s = 1,2,\cdots mn,\, t\in\mathcal{N}_s,
\end{equation}
where $I_i$ denotes the value of the $i$-th pixel of $I$, and $\mathcal{N}_s$ denotes a neighbourhood of the $s$-th pixel. Here $k = mn|\mathcal{N}_s|$ with $|\mathcal{N}_s|$ the number of elements in $\mathcal{N}_s$.

Given a target edge-histogram $\mathbf{h}$, one can perform histogram specification on $\mathbf{r}$ to obtain $\mathbf{d}$:
\begin{equation}
	\label{nonconvexmodel}
	\mathbf{r} = (r_{s_1,t_1},\cdots, {r_{s_k,t_k}})\xrightarrow[\text{specification}]{\text{histogram}}(d_{s_1,t_1},\cdots, {d_{s_k,t_k}}) =\mathbf{d}.
\end{equation}
In \cite{mignotte2012energy}, the author used a two-step procedure to accomplish \eqref{nonconvexmodel}. First, the ordering algorithm in \cite{coltuc2006exact} is applied to get a total order of ${\bf r}$. Then, with the {total order}, the entries of $\mathbf{r}$ can be re-ordered according to $\mathbf{h}$ to get ${\bf d}$.

After \eqref{nonconvexmodel}, one get the output image $X$ by solving the minimization problem
\begin{equation}
	\label{Mignotte}
	\min_X\displaystyle\sum_{s=1}^{mn}\sum_{t\in\mathcal{N}_s}\left((X_s-X_t)^2 - d^2_{s,t}\right)^2.
\end{equation}
Model \eqref{Mignotte} is solved by a conjugate gradient procedure followed by a stochastic local search. However, the non-convex nature of the model hinders the computations, and it is difficult to include additional constraints. In this paper, we propose a convex model that can include additional constraints based on different applications in edge-preserving smoothing.

Edge-preserving smoothing is a popular topic in image processing and computer graphics. Its aim is to suppress insignificant details and keep important edges intact. As an example, the input image in \Cref{fig:hist_input} contains textures on the slate and the goal of edge-preserving smoothing is to remove such textures and keep only the object boundaries as in \Cref{fig:hist_15}. Numerous methods have been introduced to perform the task. Anisotropic diffusion \cite{perona1990scale,black1998robust} performs smoothing by means of solving a non-linear PDE. Bilateral filtering \cite{tomasi1998bilateral,paris2006fast,weiss2006fast,chen2007real} is a method combining domain filters and range filters. They are widely used because of their simplicity. Optimization frameworks such as the weighted least squares (WLS) \cite{farbman2008edge} and TV regularization \cite{rudin1992nonlinear,chambolle2004algorithm} are also introduced. In WLS, a regularization term is added to minimize the horizontal and vertical gradients with corresponding smoothness weights. Recently, models based on $l_0$-gradient minimization \cite{xu2011image,cheng2014feature,storath2014jump,nguyen2015fast,pang2015improved,ono2017l_} have become popular. These models focus on the $l_0$-norm of the image gradients.

One application of edge-preserving smoothing is scan-through removal. Written or printed documents are usually subjected to various degradations. In particular, two-sided documents can be suffered from the effect of back-to-front interference, known as ``see-through'', see \Cref{background_detection}. The problem is especially severe in old documents, which is caused by the bad quality of the paper or ink-bleeding. These effects greatly reduce the readability and hinder optical character recognition. Therefore it is of great importance to remove such interference. However, physical restoration is difficult as it may damage the original contents of the documents, which is clearly undesirable as the contents may be important. Consequently, different approaches in the field of image processing are considered to restore the images digitally.

These approaches can be mainly classified into two  classes: \textit{Blind} and \textit{Non-blind} methods. Non-blind methods \cite{tonazzini2007fast,merrikh2010using,martinelli2012nonlinear,gerace2016inpainting,salerno2013nonlinear,savino2016joint,savino2016digital,sharma2001show,tonazzini2015non} require the information of both sides. These methods usually consist of two steps. First, the two sides of the images are registered. Then, the output image is computed based on the registered images. It is obvious that these methods strongly depend on the quality of registration; therefore highly accurate registration is needed. However, perfect registration is hard to achieve in practice due to numerous sources of errors including local distortions and scanning errors. Furthermore, information from the back page is not available in some occasions. Therefore, blind methods which do not assume the availability of the back page are also developed in solving the problem, see\cite{estrada2009manuscript,tonazzini2004independent,wolf2010document,sun2016blind,nishida2003correcting,tonazzini2010multichannel}.

{In this paper, we propose a convex model for applications in edge-preserving smoothing. In our work, we modify the objective function in the non-convex model in \cite{mignotte2012energy} so that we only need to solve a convex minimization problem to obtain the output. The simplicity of our model allows us to incorporate different useful constraints such as the dynamic range constraint; and the convexity of our model allows us to compute the output efficiently by Fast Iterative Shrinkage-Thresholding Algorithm (FISTA) \cite{beck2009fast} or Alternating Direction Method of Multipliers (ADMM) \cite{gabay1976dual,glowinski2008lectures}. {We introduce different edge-histograms and suitable constraints in our model,} and apply them to different imaging tasks in edge-preserving smoothing, including image abstraction, edge extraction, details exaggeration, and scan-through removal.}

The outline of the paper is as follows: \Cref{S:3} describes the proposed convex model, \Cref{S:4} presents the applications of our model with numerical results, and conclusions are then presented in \Cref{S:5}.
%%%%%%%%%%%%%%%%%%%%%%%%%%%%%%%%%%%%%%%%%%%%%%%%%%%%%%%%%%%%%%%%%%%%%%%%%%%%%%%%
%%%%%%%%%%%%%%%%%%%%%%%%%%%%%%%%%%%%%%%%%%%%%%%%%%%%%%%%%%%%%%%%%%%%%%%%%%%%%%%%
\section{Our Model}
\label{S:3}
In our model, we do not consider the edge magnitudes as in \eqref{mignotte_r}. Instead, we directly consider the image gradients and define $\mathbf{r}$ with entries
\begin{equation*}
	r_{s,t} := I_s - I_t.
\end{equation*}
Similar to \cite{mignotte2012energy}, our model consists of two parts. First, given $\mathbf{r}$, we perform histogram specification on $\mathbf{r}$ to obtain $\mathbf{d}$ as in \eqref{nonconvexmodel}. In the second part, we solve a minimization problem to obtain the output {$X$}. Instead of solving the non-convex model \eqref{Mignotte}, we propose a convex model. In \Cref{pro_algo}, we present our convex model and its solvers. In \Cref{histconstruct,ConvexSetC,iters}, we apply our model to specific applications in edge-preserving smoothing.

In the following discussions, we consider only grayscale images. For colored images, we apply our method to R, G, B channels separately.
%%%%%%%%%%%%%%%%%%%%%%%%%%%%%%%%%%%%%%%%%%%%%%%%%%%%%%%%%%%%%%%%%%%%%%%%%%%%%%%%
\subsection{{Proposed convex model}}
\label{pro_algo}
Instead of \eqref{Mignotte}, we propose the convex model
\begin{equation}
	\label{Proposed_algo_ori}
	\min_X\displaystyle\sum_{s=1}^{mn}\sum_{t\in\mathcal{N}_s}\left|(X_s-X_t) - d_{s,t}\right|^p + \iota_\mathbf{C}(X),
\end{equation}
where $p= 1$ or $2$, $\mathbf{C}$ is a convex set to be discussed in \Cref{ConvexSetC}, and $\iota_\mathbf{C}$ denotes the indicator function of $\mathbf{C}$. The choice of $p$ and $\mathbf{C}$ depends on applications. Let $\mathbf{x}$ be a vector such that its $s$-th entry is $X_s$. Then we can rewrite \eqref{Proposed_algo_ori} as
\begin{equation}
	\label{Proposed_algo}
	\min_\mathbf{x} ||G\mathbf{x} - \mathbf{d}||_p^p + \iota_\mathbf{C}(\mathbf{x}).
\end{equation}
In our tests, we use $\mathcal{N}_s = \{s_v,\,s_h\}$, where $s_v$ and $s_h$ denote the pixel above and at the left of the $s$-th pixel. Hence we can write $G = (G_h,G_v)^T$, where $G_h$, $G_v$ are the horizontal and vertical backward difference operators. We use periodic boundary condition for pixels outside the boundaries, see \cite[p.~258]{gonzales1992digital}.

For $p=2$, model \eqref{Proposed_algo} can be solved by FISTA \cite{beck2009fast}. For $p=1$, we rewrite \eqref{Proposed_algo} as 
\begin{equation*}
	\begin{aligned}
		&\min_{\mathbf{x},\mathbf{y}} \|\mathbf{y} - \mathbf{d}\|_1 + \iota_\mathbf{C}(\mathbf{x})\\
		&\,\,\text{s.t. }G\mathbf{x}=\mathbf{y} ,
	\end{aligned}
\end{equation*}
which can be solved by ADMM \cite{gabay1976dual,glowinski2008lectures}.
%%%%%%%%%%%%%%%%%%%%%%%%%%%%%%%%%%%%%%%%%%%%%%%%%%%%%%%%%%%%%%%%%%%%%%%%%%%%%%%%
\subsection{Construction of target edge-histogram}
\label{histconstruct}
\begin{figure}[htb]
  	\centering
  	\subfloat[Input]{\includegraphics[width=0.3\textwidth,clip]{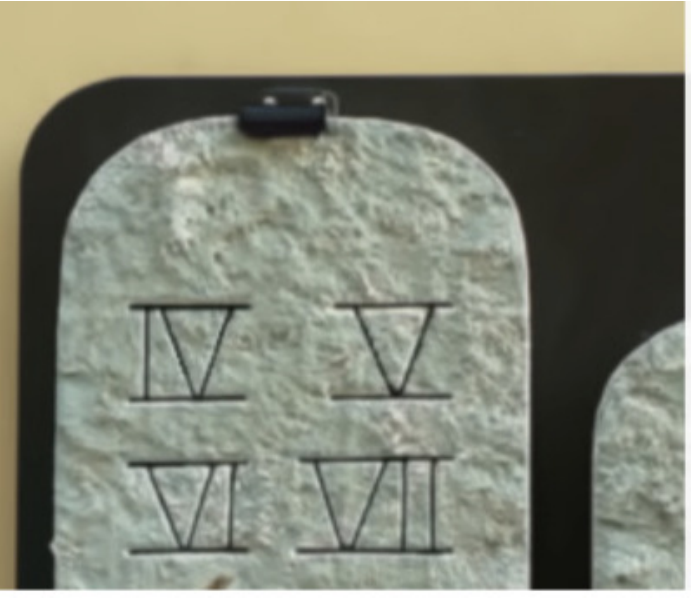}\label{fig:g_input}}\,
  	\subfloat[$|\text{horizontal gradient}|$]{\includegraphics[width=0.3\textwidth,clip]{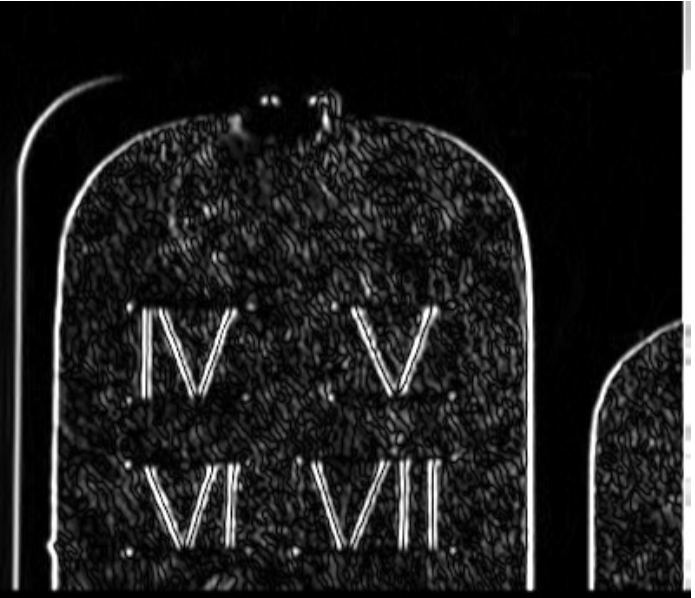}\label{fig:h_gradient}}\,
  	\subfloat[$|\text{vertical gradient}|$]{\includegraphics[width=0.3\textwidth,clip]{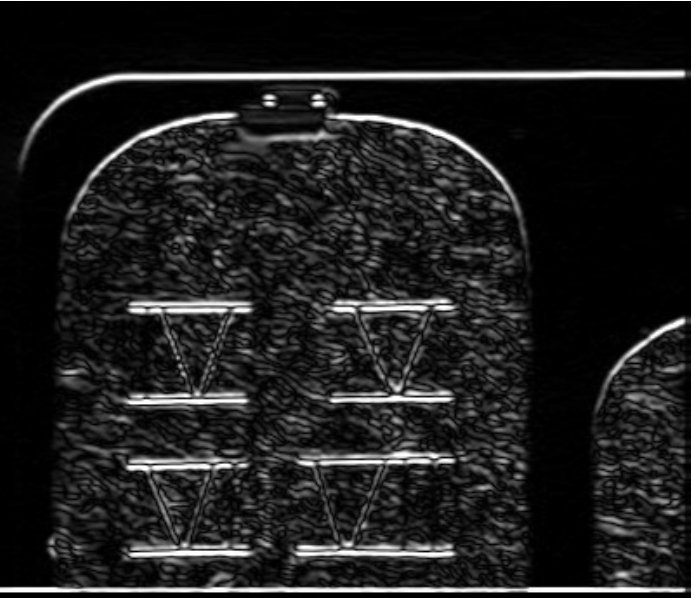}\label{fig:v_gradient}}\,
  	\caption{Given image and its horizontal and vertical gradients in absolute values.}
 	\label{assumption}
\end{figure}

\begin{figure}[htb]
	\centering
  	\subfloat[Histogram of gradient $y_{s,t}$]{\includegraphics[width=0.45\textwidth,clip]{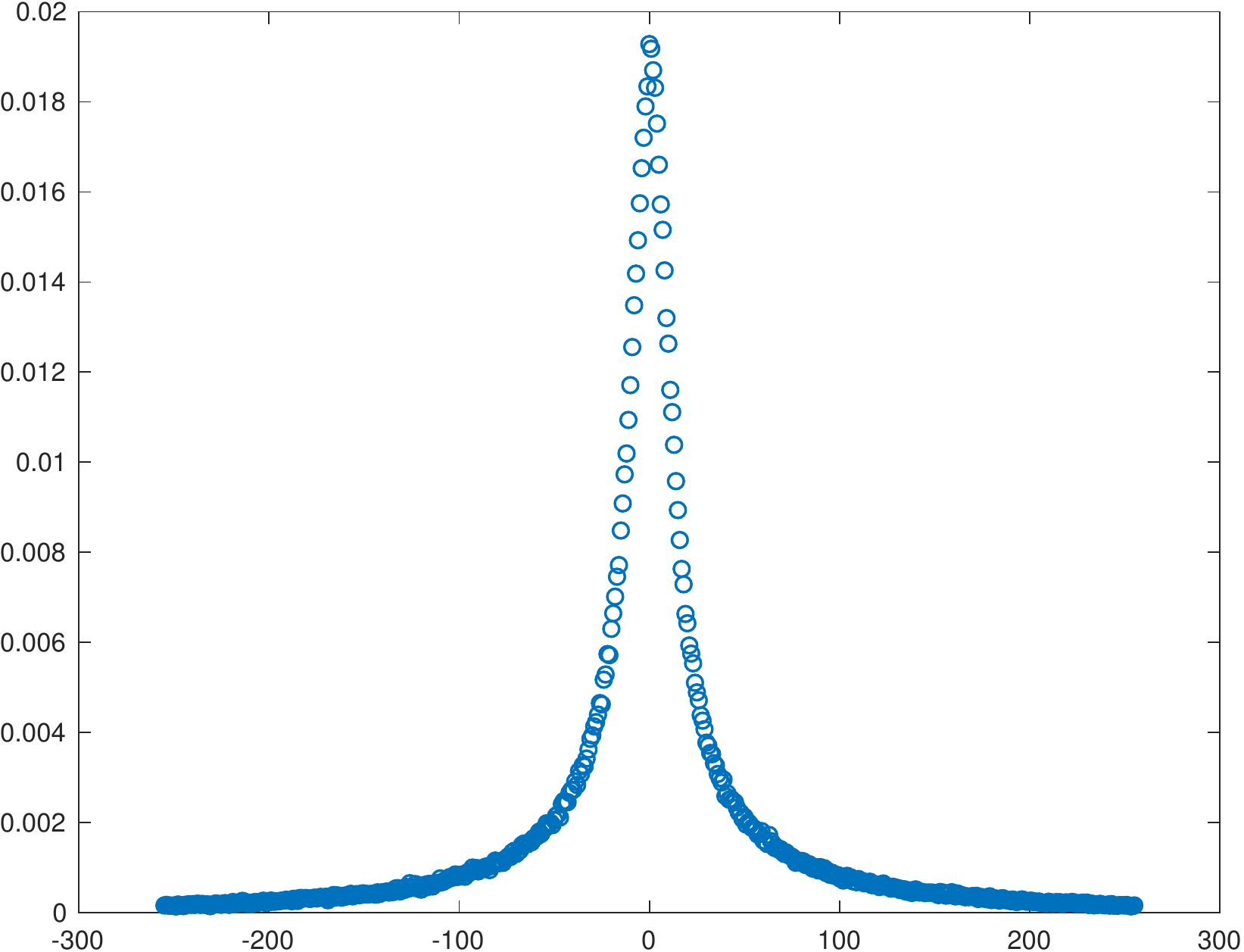}\label{fig:TargetHist_in}}\,
  	\subfloat[Histogram of thresholded gradient $z_{s,t}$]{\includegraphics[width=0.45\textwidth,clip]{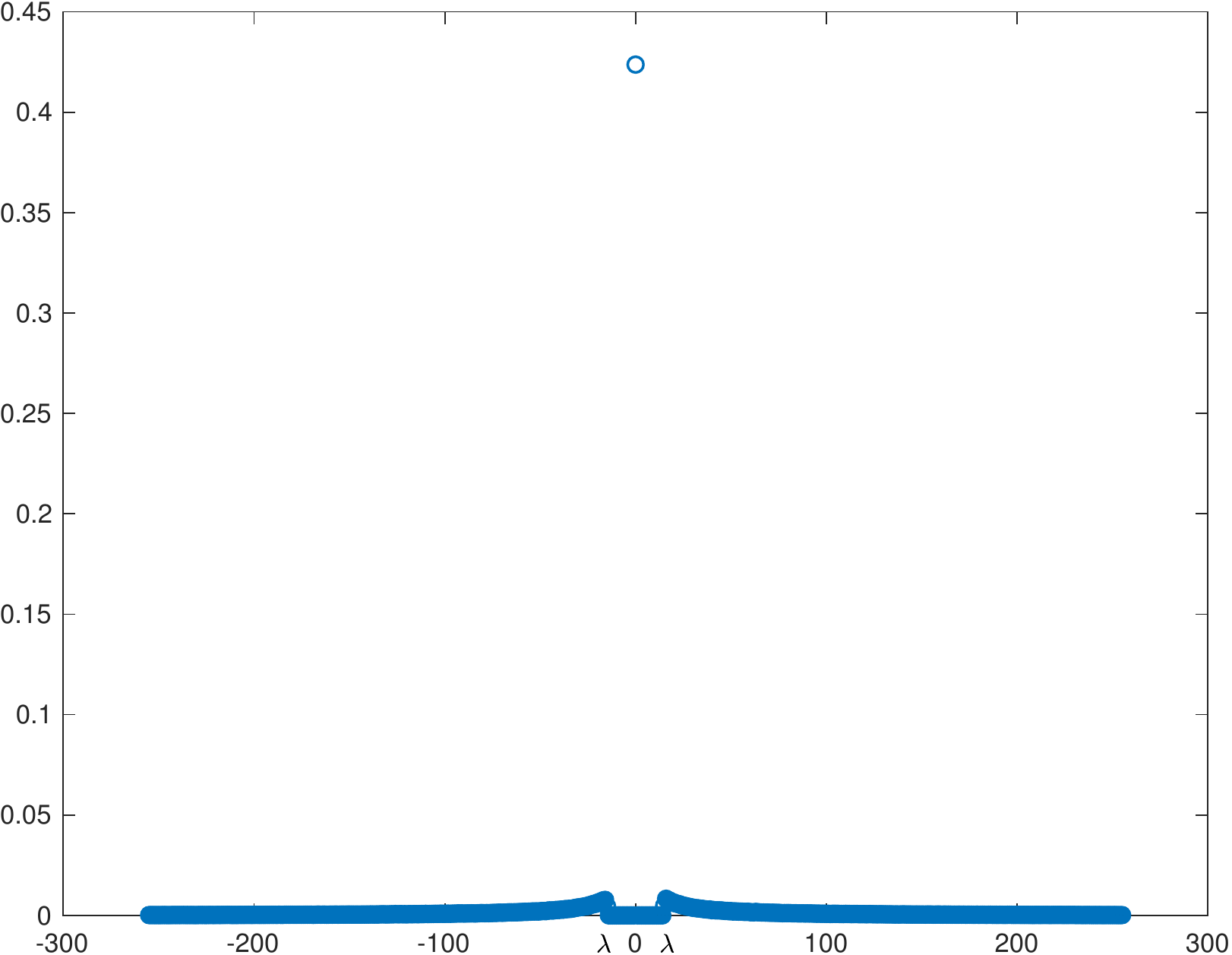}\label{fig:TargetHist_out}}
  	\caption{Construction of the target edge-histogram.}
  	\label{TargetHist}
\end{figure}

For the applications we considered in this paper, one objective is to remove  textures in the images where
their edges have small magnitude.
As an example, the textures on the slate in \Cref{fig:g_input} produce smaller edge magnitudes compared to the boundaries of the slate and the letters on the slate, see \Cref{fig:h_gradient,fig:v_gradient}. To eliminate those textures, we could set the values of edges with small magnitude to zero. Hence, in this paper, we propose to use edge-histograms similar to that shown in \Cref{fig:TargetHist_out} as target edge-histogram which is obtained by thresholding the input edge-histogram in \Cref{fig:TargetHist_in}. In particular, the target edge-histogram is dependent on the input image. We remark here that it is not uncommon to construct the target histogram based on the input. For example, such construction is used in image segmentation \cite{thomas2008image}.

\begin{figure}[t]
  	\centering
  	\subfloat[Input]{\includegraphics[width=0.3\textwidth,clip]{hist_input}\label{fig:hist_input}}\,
  	\subfloat[$\lambda=5$]{\includegraphics[width=0.3\textwidth,clip]{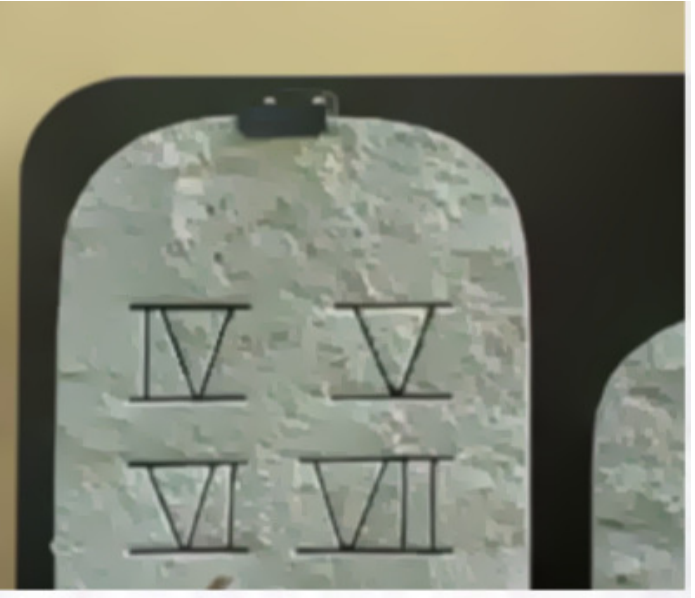}\label{fig:hist_5}}\,
  	\subfloat[$\lambda=15$]{\includegraphics[width=0.3\textwidth,clip]{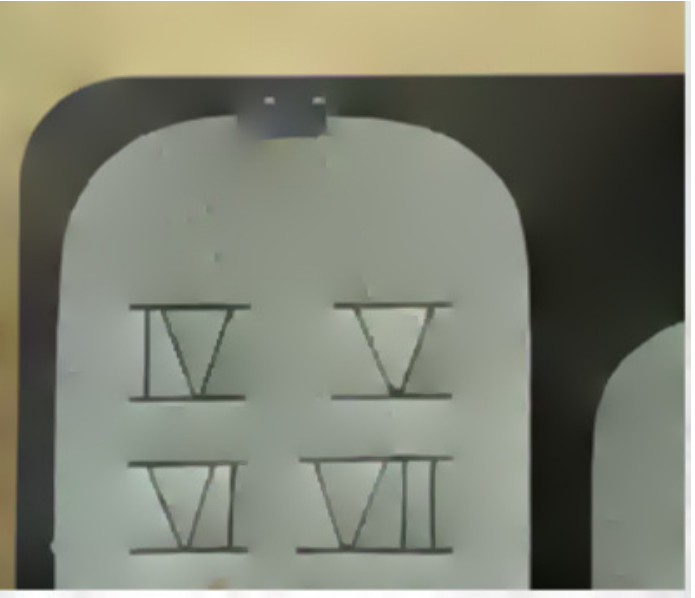}\label{fig:hist_15}}
  	\caption{Output of {\eqref{Proposed_algo}} with different $\lambda$.}
  	\label{hist_comp}
\end{figure}

Since we are just thresholding the edges with small values to zero, the edge-histogram specification \eqref{nonconvexmodel} can be done easily as follows. Given any input image $Y$, we first compute its gradients $y_{s,t} = Y_s -Y_t$. Then we set
\begin{equation}
	\label{threshold}
	z_{s,t} = 
	\begin{cases}
		y_{s,t} \quad &\text{if } |y_{s,t}| \geq \lambda,\\
		0 \quad &\text{otherwise.}
	\end{cases}	
\end{equation}
The thresholded $z_{s,t}$, where its histogram is shown in \Cref{fig:TargetHist_out}, will be used as the vector ${\bf d}$ in \eqref{Proposed_algo} to obtain the output ${\bf x}$. It is obvious that different $\lambda$ gives different outputs, see \Cref{hist_comp}. We see that the smoothness of the output increases with $\lambda$.

%%%%%%%%%%%%%%%%%%%%%%%%%%%%%%%%%%%%%%%%%%%%%%%%%%%%%%%%%%%%%%%%%%%%%%%%%%%%%%%%
\subsection{{Gaussian smoothing and iterations}}
\label{iters}
Strong textures can produce edges with large magnitude that cannot be eliminated using a thresholded edge-histogram as in \Cref{fig:TargetHist_out}. To suppress them, the input image $I$ will first pass through a Gaussian filter with standard deviation $\sigma$ to get the initial guess $X^{(0)}$. Larger $\sigma$ will have a greater effect in suppressing strong textures, but at the same time blur the image. Hence, $\sigma$ should be chosen small enough so that the Gaussian-filtered image is visually equal to $I$. Let $X^{(0)}$ be the Gaussian-filtered image. Whenever such suppression is unnecessary, we set $\sigma = 0$ and hence $X^{(0)} = I$.

\begin{figure}[htb]
  	\centering
  	\subfloat[Input image $I$]{\includegraphics[width=0.3\textwidth,clip]{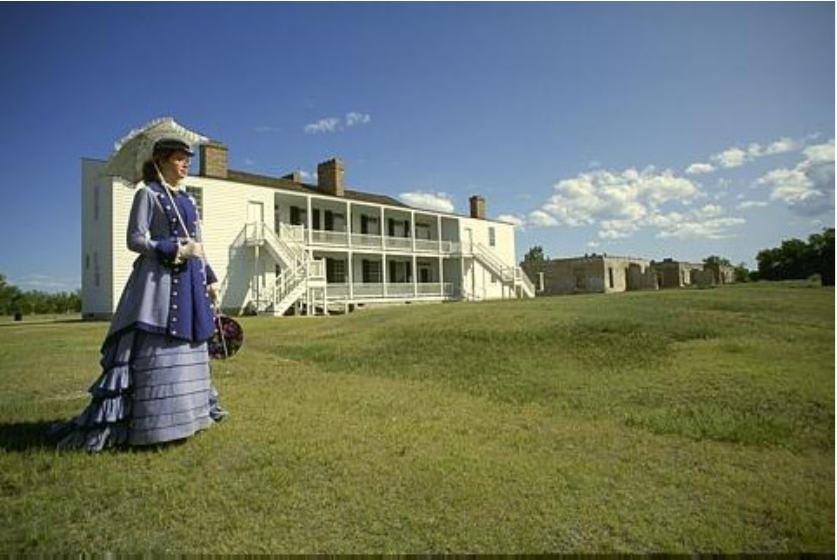}\label{fig:iter_input}}\,
  	\subfloat[After one  iteration]{\includegraphics[width=0.3\textwidth,clip]{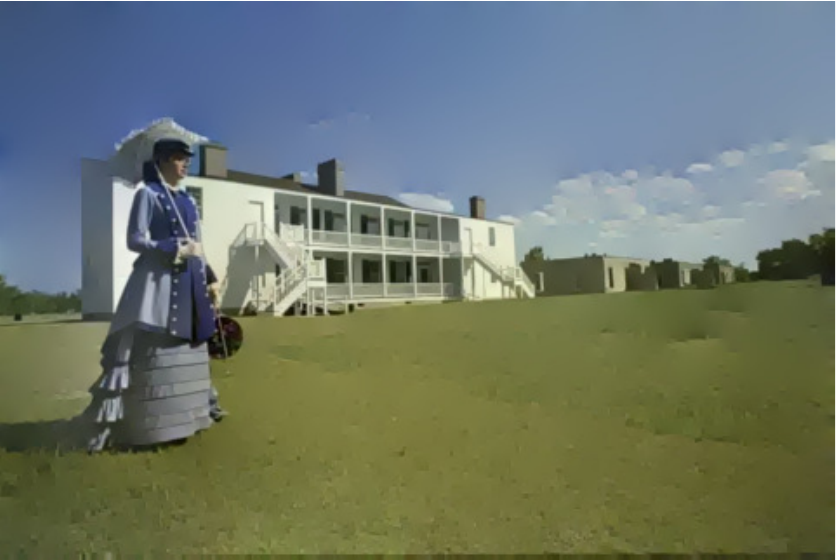}\label{fig:1_iter}}\,
  	\subfloat[After three iterations]{\includegraphics[width=0.3\textwidth,clip]{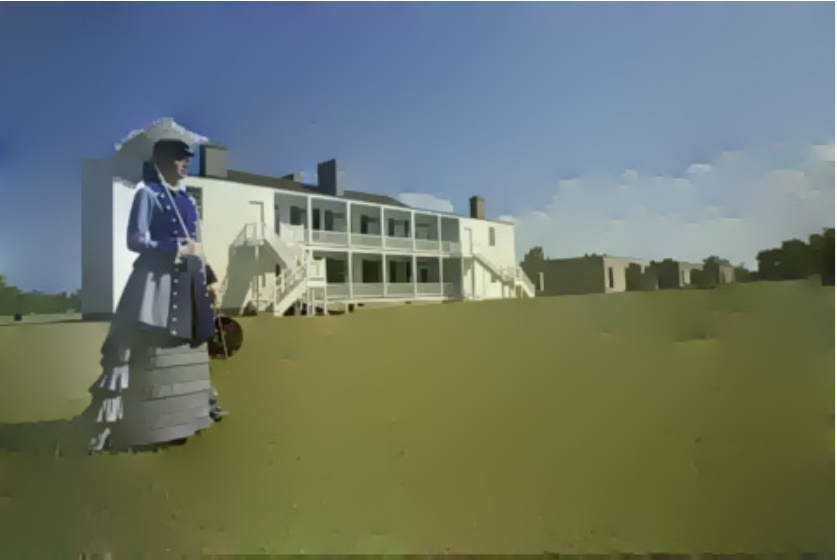}\label{fig:3_iter}}
  	\caption{Output of solving {\eqref{Proposed_algo}} repeatedly. Here $\lambda = 15, \sigma = 0.6$.}
  	\label{iterations}
\end{figure}

As mentioned in \Cref{histconstruct}, one of our objectives is to map small edges to zero. This can be done by changing the $\lambda$ in the thresholded edge-histogram or by solving {\eqref{Proposed_algo}} repeatedly. More specifically, given $X^{(0)}$, we construct $\mathbf{d}$ using \eqref{threshold} and solve \eqref{Proposed_algo} to obtain $X^{(1)}$. Then we repeat the process to obtain $X^{(2)}$ and so on. \Cref{iterations} shows a comparison; while we see in \Cref{fig:1_iter} that the result after one iteration still contains textures in the grasses, almost all of them are removed after three iterations, see \Cref{fig:3_iter}.
%%%%%%%%%%%%%%%%%%%%%%%%%%%%%%%%%%%%%%%%%%%%%%%%%%%%%%%%%%%%%%%%%%%%%%%%%%%%%%%%
\subsection{Convex set $\mathbf{C}$}
\label{ConvexSetC}
The model \eqref{Mignotte} does not consider the dynamic range constraint
\begin{equation}
	\label{dynamic_range}
	I_s \in[0,255],\,\,\forall s=1,2,\cdots,mn.
\end{equation}
For example, consider the case when one defines $\mathbf{h}$ such that every pixel of $\mathbf{r}$ is doubled. In the absence of (\ref{dynamic_range}), it is easy to get an exact solution {$X$} of \eqref{Mignotte} if any one of the pixel values is given.  However, there is no guarantee that the pixel values of $X$ lies within $[0,255]$. Therefore, when $X$  is converted back to the desired dynamic range, either by stretching or clipping, the edge-histogram is no longer preserved. To avoid this, it is better to include the dynamic range constraint in the objective function. Therefore, we use the following constraint in all our applications:
\begin{equation} 
	\label{C1}
	\mathbf{C} = \{\mathbf{x}:x_i\in[0,255],\,\,\forall i\}.
\end{equation}

In scan-through removal, we assume the background in books and articles have a lighter intensity than the ink in all color channels. Therefore, in addition to the dynamic range constraint, we also keep the value of the background pixels unchanged. Hence, we set
\begin{equation}
	\label{C_scan}
	\mathbf{C} = \{\mathbf{x}:x_i\in[0,255],\,\,\forall i \text{ and }x_i = X_{i}^{(0)} \text{ if } x_i \geq \alpha\},
\end{equation}
where $\alpha$ is the approximate intensity of the background to be defined in \Cref{S:scanthrough}.
%%%%%%%%%%%%%%%%%%%%%%%%%%%%%%%%%%%%%%%%%%%%%%%%%%%%%%%%%%%%%%%%%%%%%%%%%%%%%%%%
%%%%%%%%%%%%%%%%%%%%%%%%%%%%%%%%%%%%%%%%%%%%%%%%%%%%%%%%%%%%%%%%%%%%%%%%%%%%%%%%
\section{Applications and Comparisons}
\label{S:4}
Edge-preserving smoothing includes many different applications. In this section, we consider four applications, namely image abstraction, edge extraction, details exaggeration, and scan-through removal. For the first three applications, we use $p=2$ in \eqref{Proposed_algo} and solve it by FISTA with the input image as initial guess. We compare with four existing methods: {bilateral filtering} \cite{paris2006fast}, {weighted-least square} \cite{farbman2008edge}, {$l_0$-smoothing} \cite{xu2011image}, and {$l_0$-projection} \cite{ono2017l_}. For the scan-through removal, we use $p=1$ in \eqref{Proposed_algo} and solve it by ADMM with the input image and $\mathbf{d}$ as the initial guesses. We compare with one blind method \cite{nishida2003correcting} and three non-blind methods \cite{martinelli2012nonlinear, tonazzini2010multichannel, hyvarinen1999fast}.
In all applications, the number of iterations is fixed at $3$. The values $\lambda$ and $\sigma$ vary for different images and will be stated separately.

For the tests below, we select the parameters which give the output image with the best visual quality. Some of the comparison results are obtained directly from the authors' work and some are done by ourselves. For the results done by ourselves, we list out the parameters we have used.
%%%%%%%%%%%%%%%%%%%%%%%%%%%%%%%%%%%%%%%%%%%%%%%%%%%%%%%%%%%%%%%%%%%%%%%%%%%%%%%%
\subsection{Image abstraction}
\begin{figure}[htb]
  \centering
  \subfloat[Input]{\includegraphics[width=0.2\textwidth,clip]{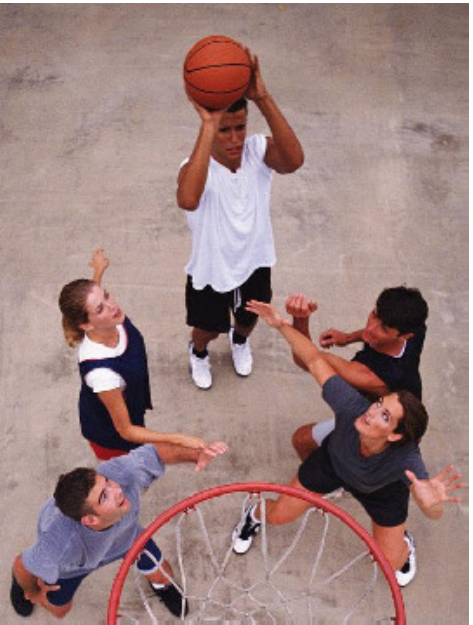}\label{fig:abs1_input}}\,
  \subfloat[][\centering Bilateral \cite{paris2006fast}

  $\sigma_s = 8,\,\sigma_r = 26$]{\includegraphics[width=0.2\textwidth,clip]{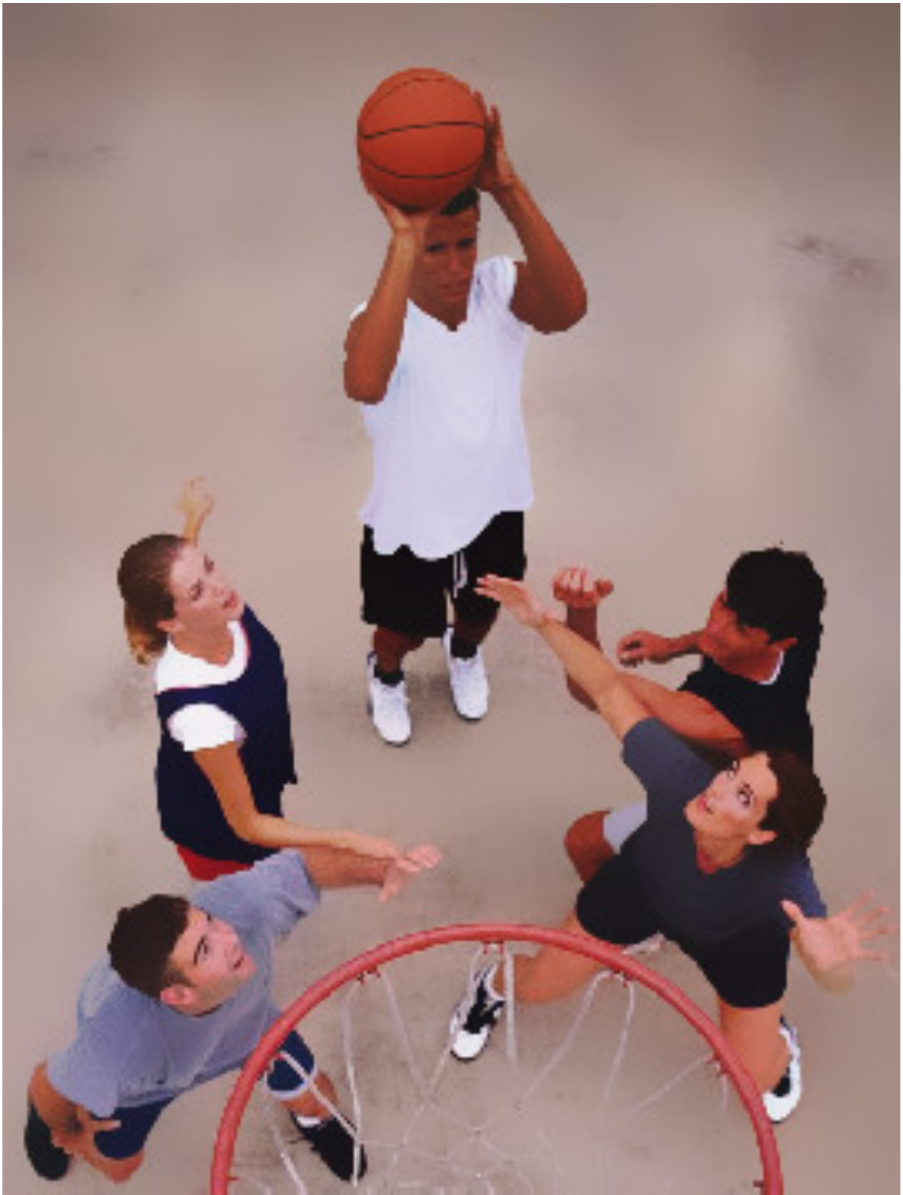}\label{fig:abs1_Bilateral}}\,
  \subfloat[][\centering WLS \cite{farbman2008edge}

  $\alpha = 1.5,\,\lambda = 0.5$]{\includegraphics[width=0.2\textwidth,clip]{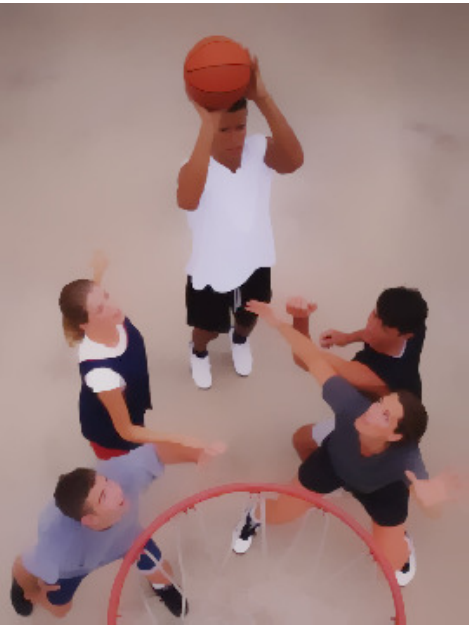}\label{fig:abs1_WLS}}\\
  \subfloat[$l_0$-smoothing \cite{xu2011image}]{\includegraphics[width=0.2\textwidth,clip]{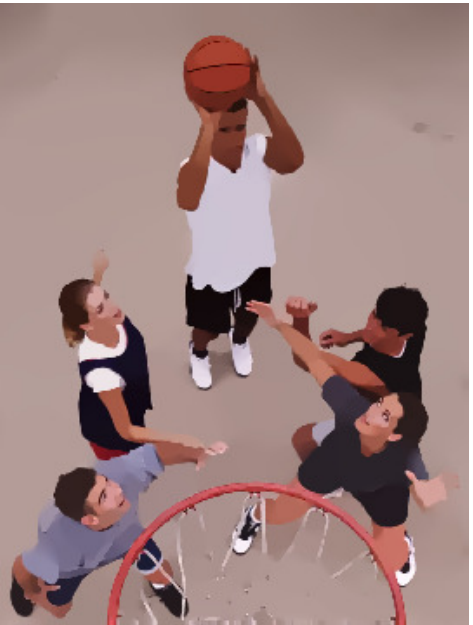}\label{fig:abs1_Jia}}\,
  \subfloat[][\centering$l_0$-projection \cite{ono2017l_}

  $\alpha = 21749$]{\includegraphics[width=0.2\textwidth,clip]{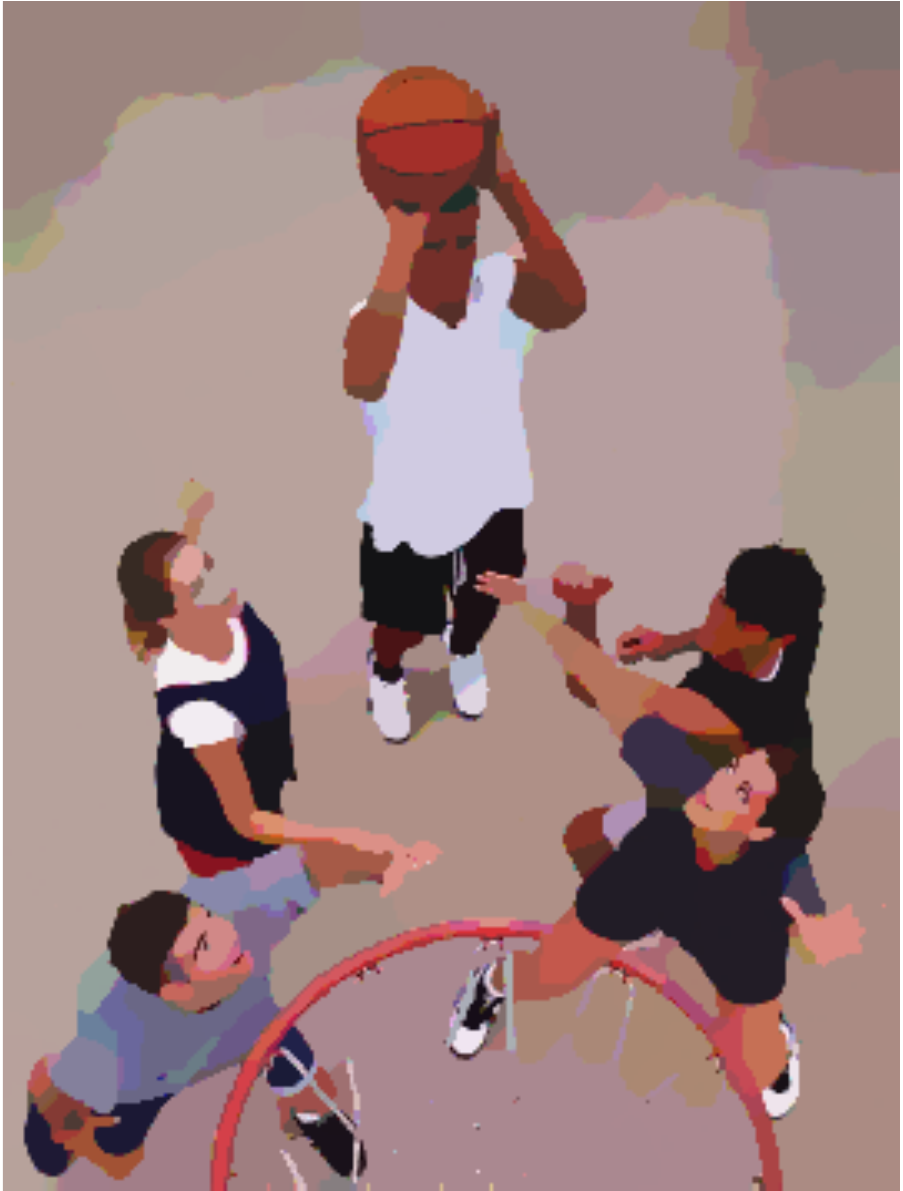}\label{fig:abs1_l0proj}}\,
  \subfloat[][\centering Ours

  \centering$\lambda=15$, $\sigma=0$]{\includegraphics[width=0.2\textwidth,clip]{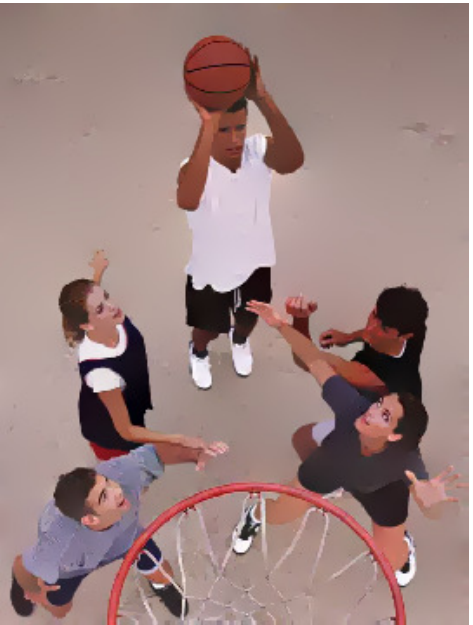}\label{fig:abs1_output}}
  \caption{Comparison of our method with other methods in image abstraction.}
  \label{abs}
\end{figure}

The goal of image abstraction is to remove textures and fine details so that the output looks un-photorealistic. This can be done by solving \eqref{Proposed_algo} with constraint (\ref{C1}). As shown in \Cref{abs}, the textures of the objects in the photorealistic input image in \Cref{fig:abs1_input} is removed and our output in \Cref{fig:abs1_output} becomes un-photorealistic. We see that our model successfully eliminate almost all object textures and keep the object boundaries intact. As we see in \Cref{fig:abs1_output}, the details in the basketball net in our output are kept intact, while it disappears in the output of other models.
%%%%%%%%%%%%%%%%%%%%%%%%%%%%%%%%%%%%%%%%%%%%%%%%%%%%%%%%%%%%%%%%%%%%%%%%%%%%%%%%
\subsection{Edge extraction}
\begin{figure}[htb]
  \centering
  \subfloat[Input]{\includegraphics[width=0.3\textwidth,clip]{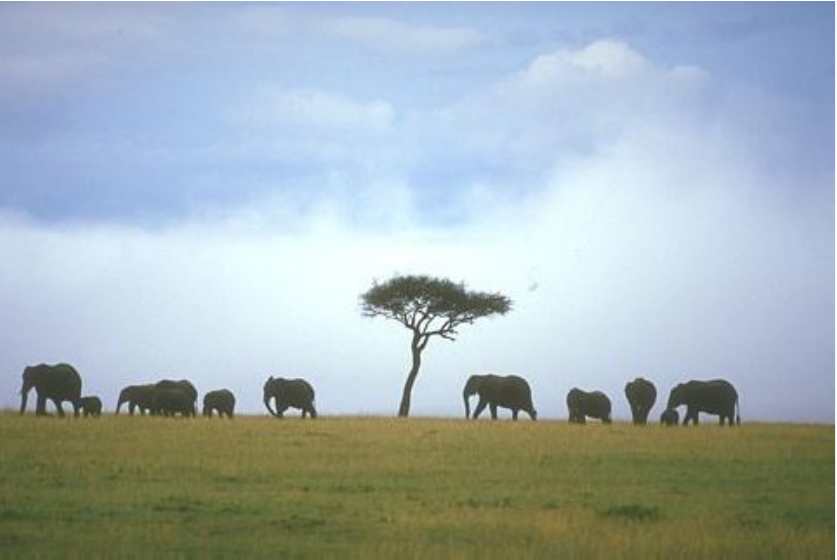}\label{fig:eg1_input}}\,
  \subfloat[][\centering Bilateral \cite{paris2006fast}

  $\sigma_s = 5$, $\sigma_r = 46$]{\includegraphics[width=0.3\textwidth,clip]{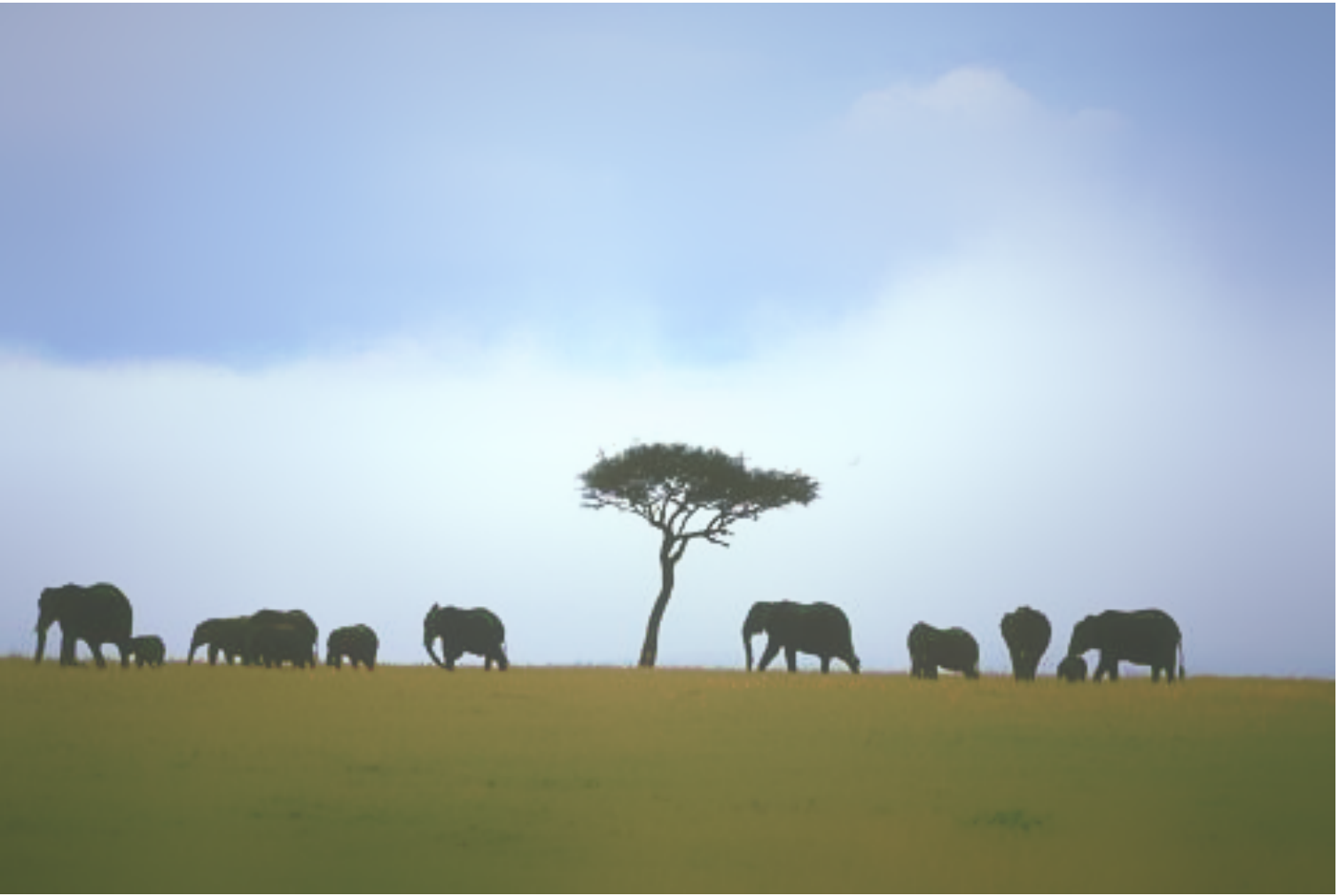}\label{fig:eg1_bilateral}}\,
    \subfloat[WLS \cite{farbman2008edge}, $\alpha = 2$, $\lambda = 2$]{\includegraphics[width=0.3\textwidth,clip]{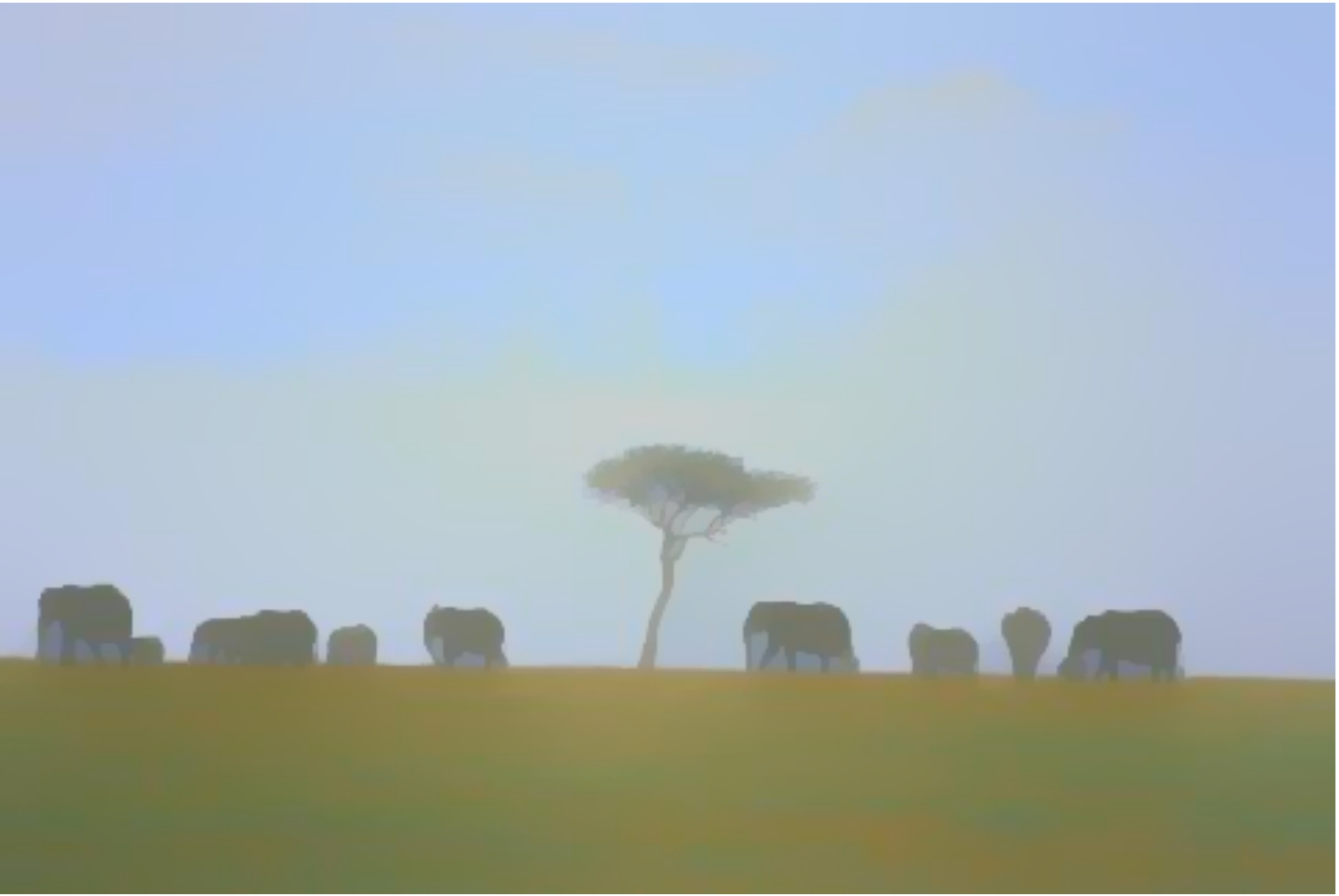}\label{fig:eg1_WLS}}\\
  \subfloat[$l_0$-smoothing \cite{xu2011image}]{\includegraphics[width=0.3\textwidth,clip]{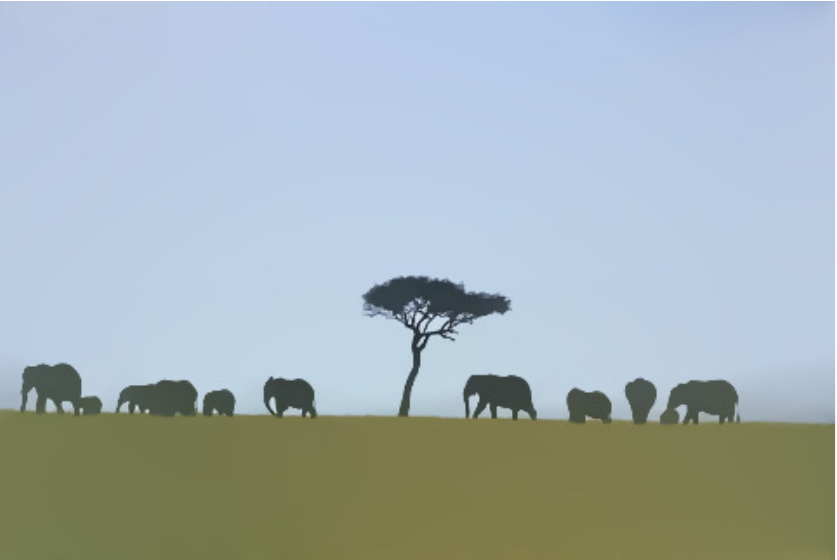}\label{fig:eg1_jia}}\,
  \subfloat[$l_0$-projection \cite{ono2017l_}, $\alpha = 9264$]{\includegraphics[width=0.3\textwidth,clip]{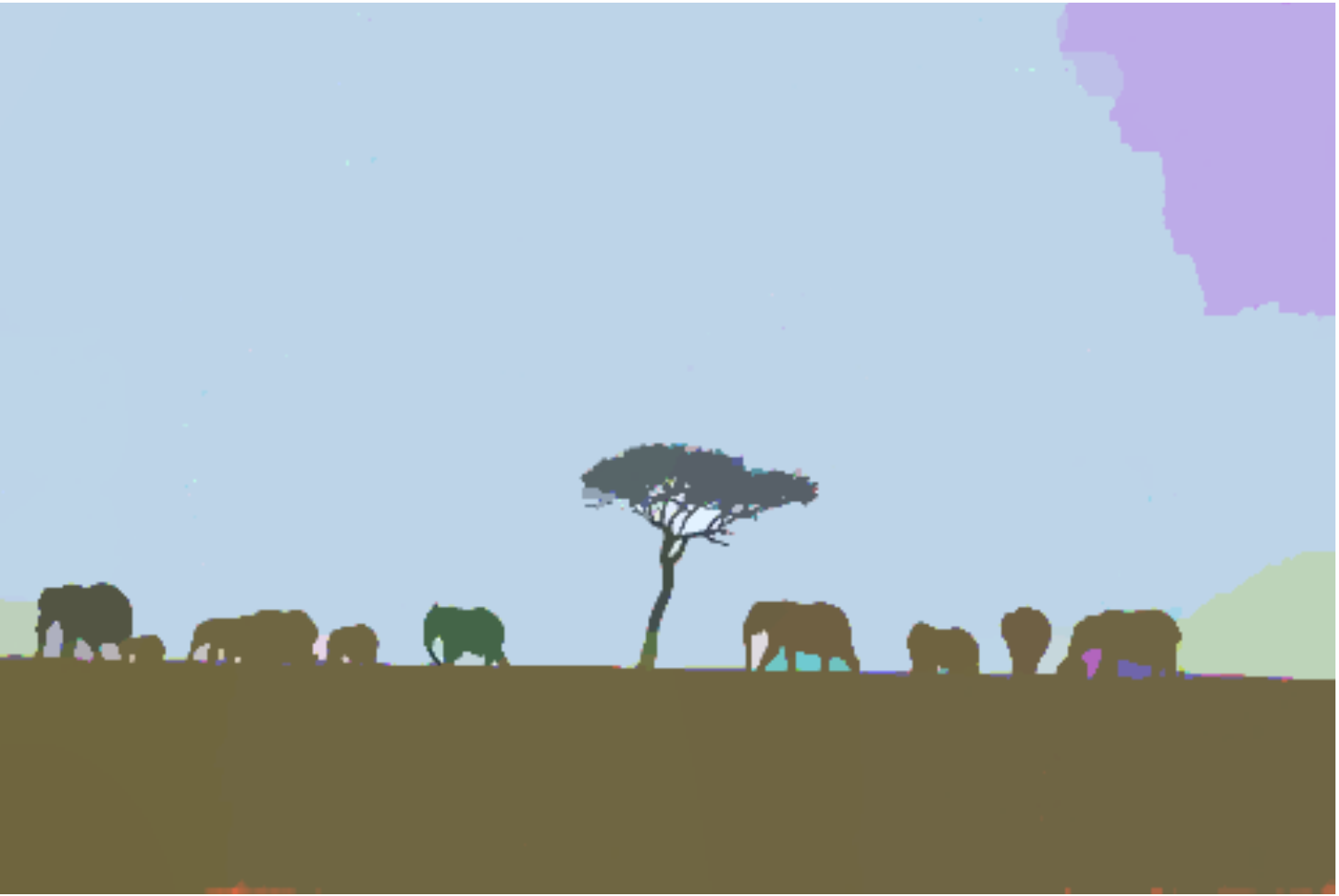}\label{fig:eg1_l0proj}}\,
  \subfloat[Ours, $\lambda=10$, $\sigma = 0.7$]{\includegraphics[width=0.3\textwidth,clip]{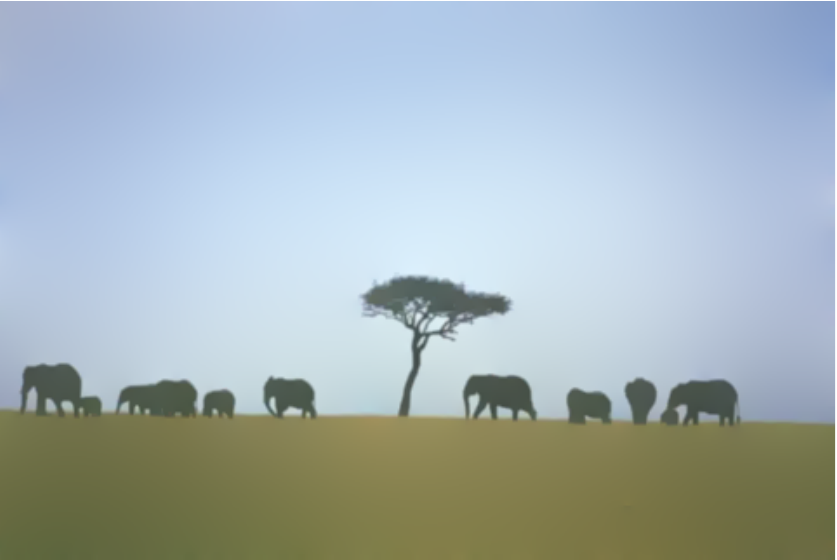}\label{fig:eg1_output}}\\
  \caption{Comparison of our method with other methods in edge extraction.}
  \label{edges1}
\end{figure}

\begin{figure}[htb]
  \centering
  \subfloat[Input]{\includegraphics[width=0.3\textwidth,clip]{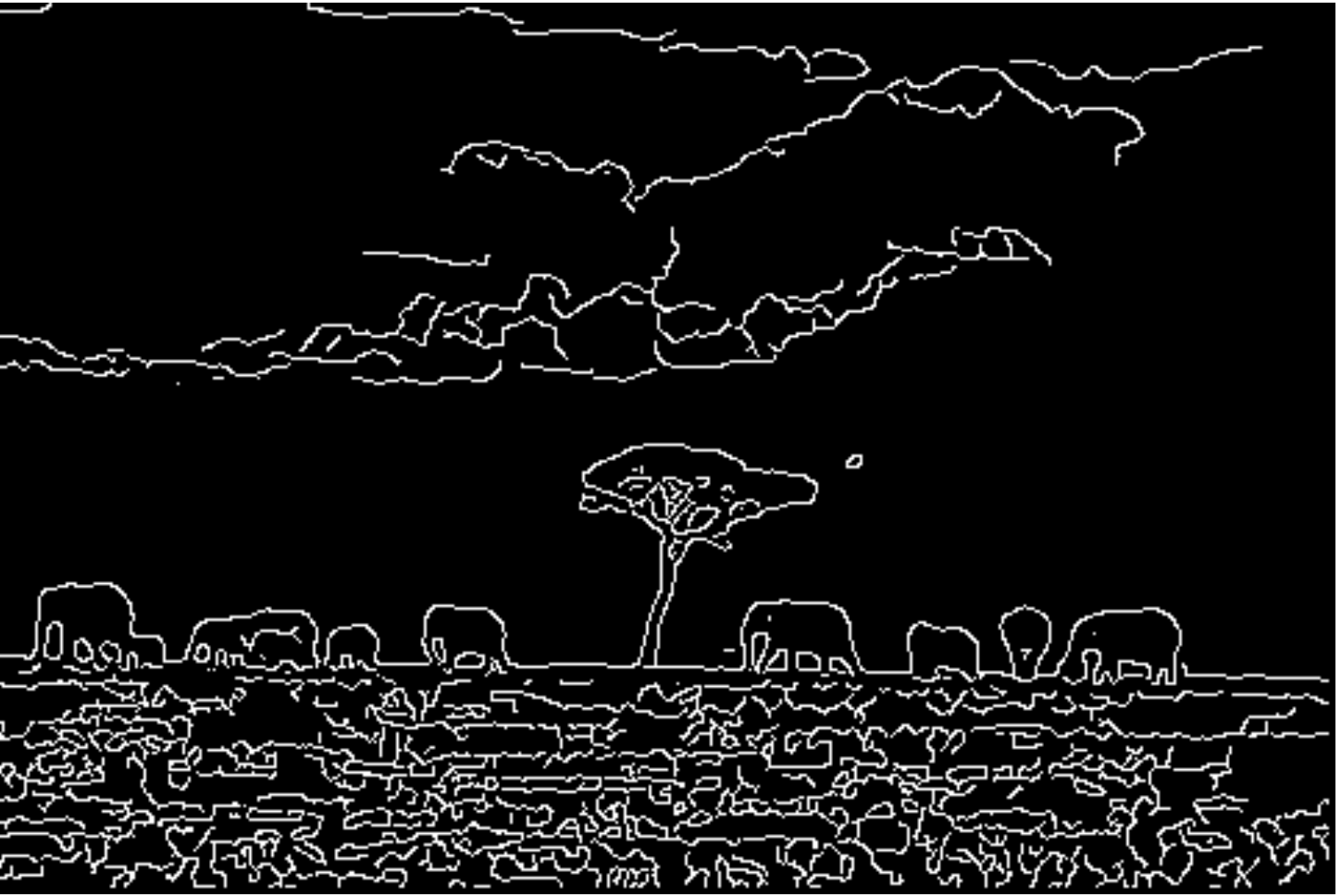}\label{fig:eg1_input_canny}}\,
  \subfloat[Bilateral \cite{paris2006fast}]{\includegraphics[width=0.3\textwidth,clip]{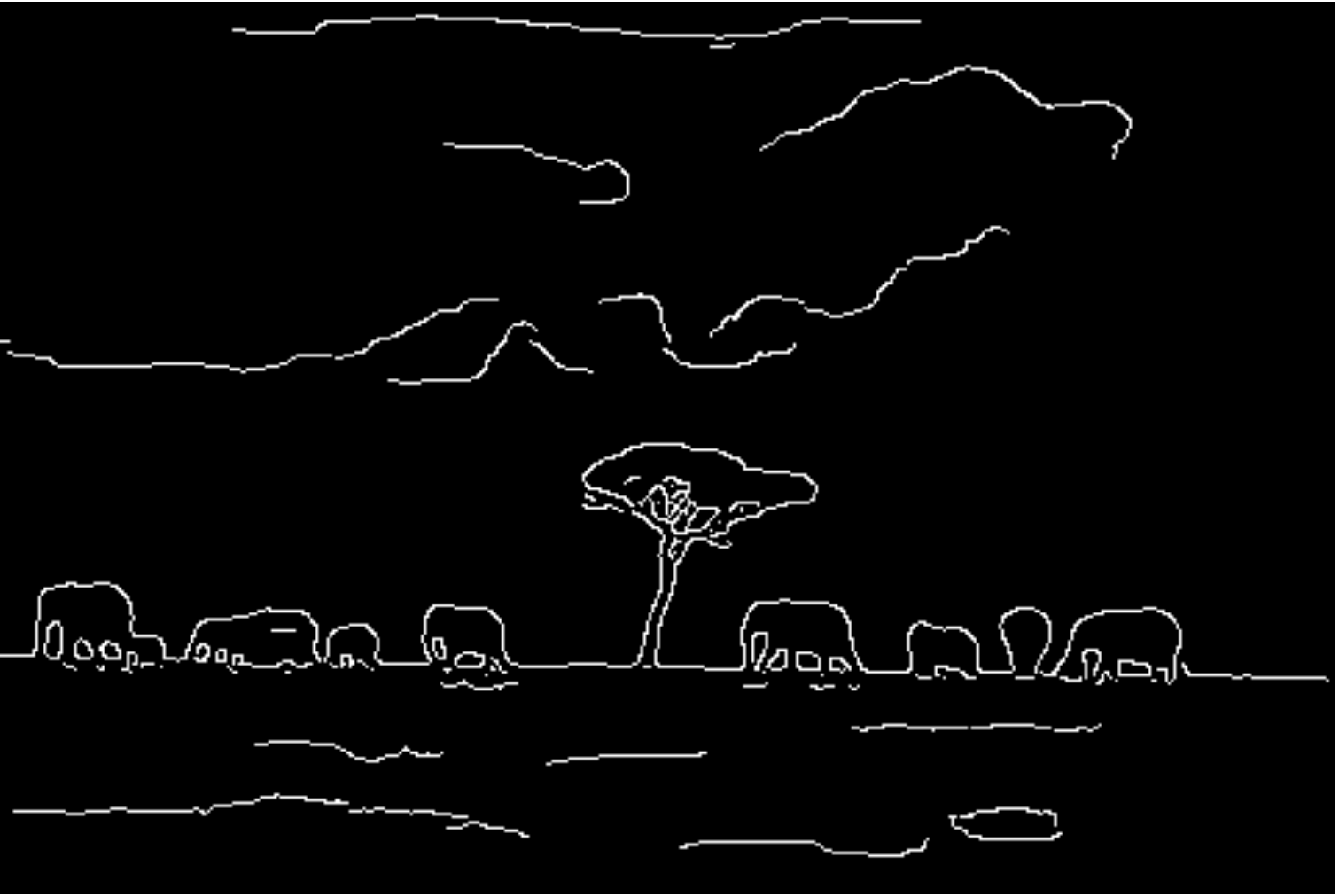}\label{fig:eg1_bilateral_canny}}\,
  \subfloat[WLS \cite{farbman2008edge}]{\includegraphics[width=0.3\textwidth,clip]{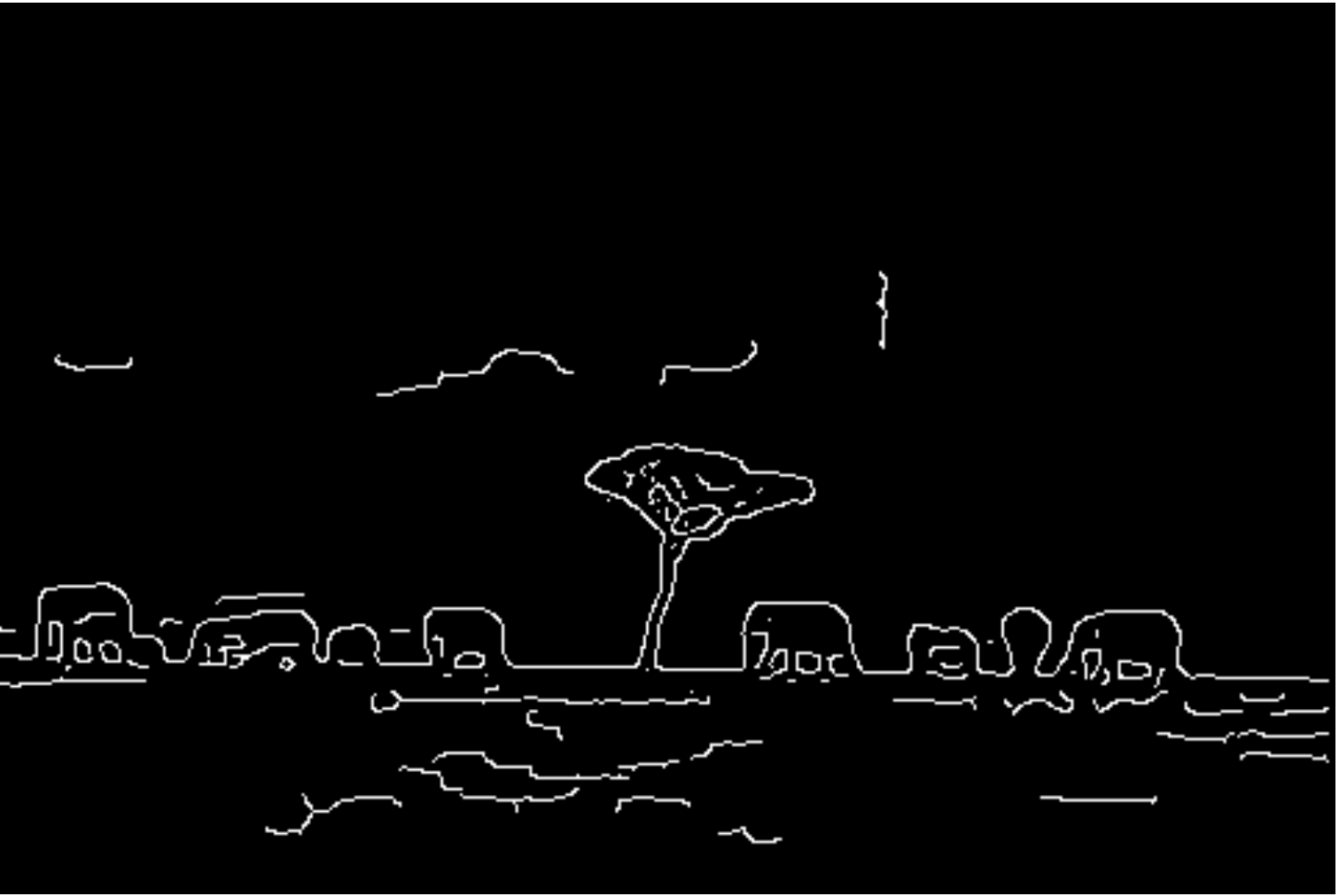}\label{fig:eg1_WLS_canny}}\\
  \subfloat[$l_0$-smoothing \cite{xu2011image}]{\includegraphics[width=0.3\textwidth,clip]{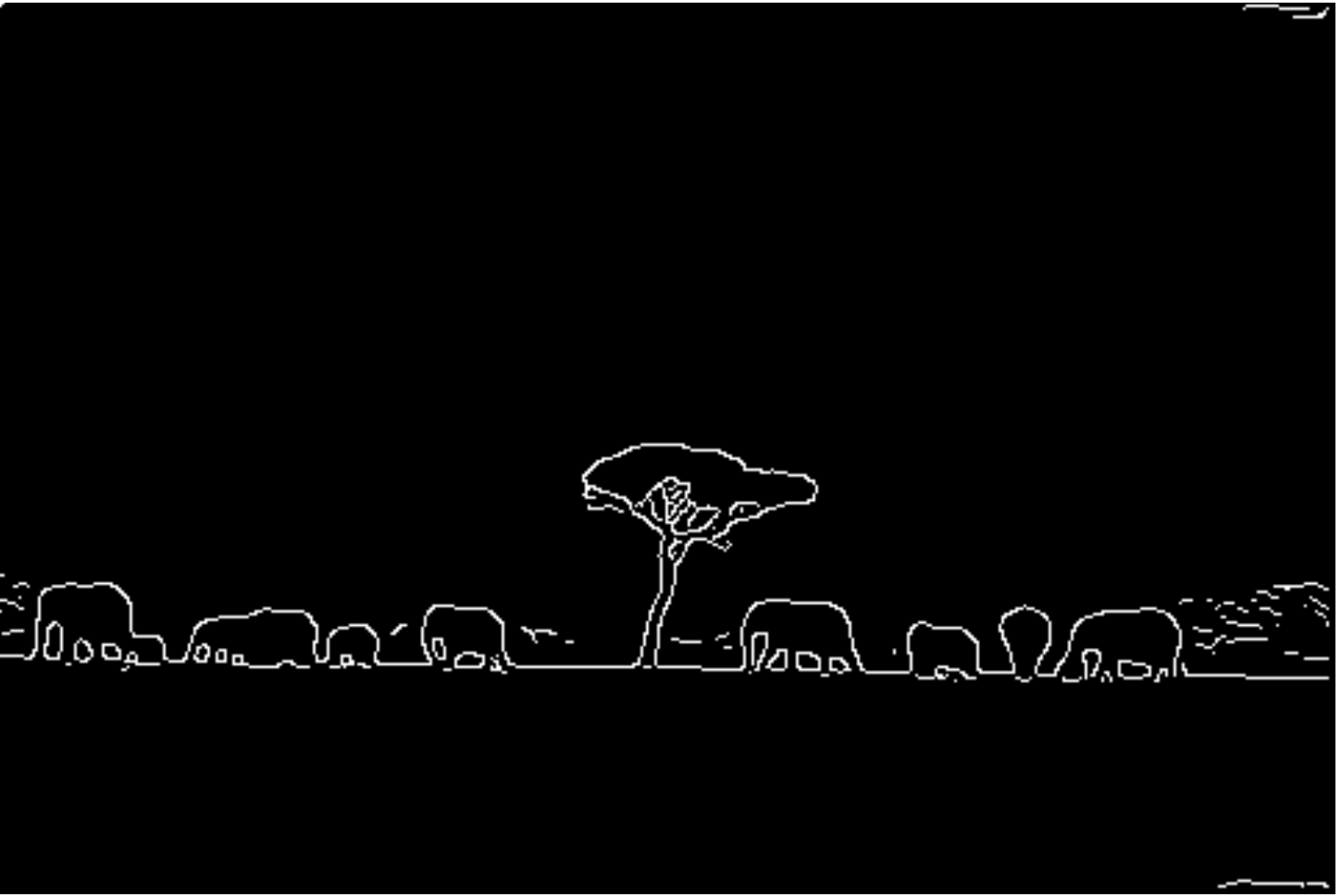}\label{fig:eg1_jia_canny}}\,
  \subfloat[$l_0$-projection \cite{ono2017l_}]{\includegraphics[width=0.3\textwidth,clip]{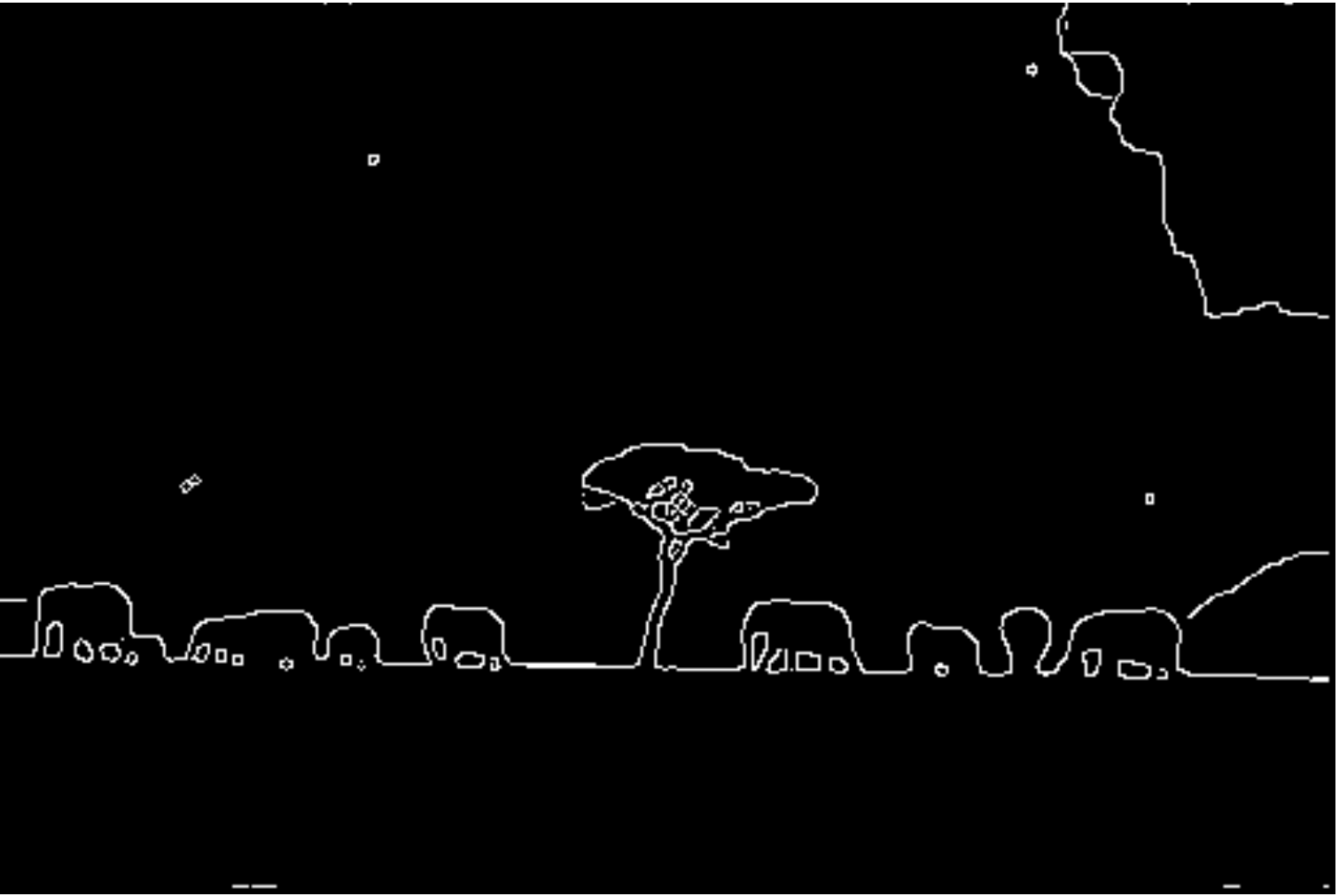}\label{fig:eg1_l0proj_canny}}\,
  \subfloat[Ours, $\lambda=10$, $\sigma = 0.7$]{\includegraphics[width=0.3\textwidth,clip]{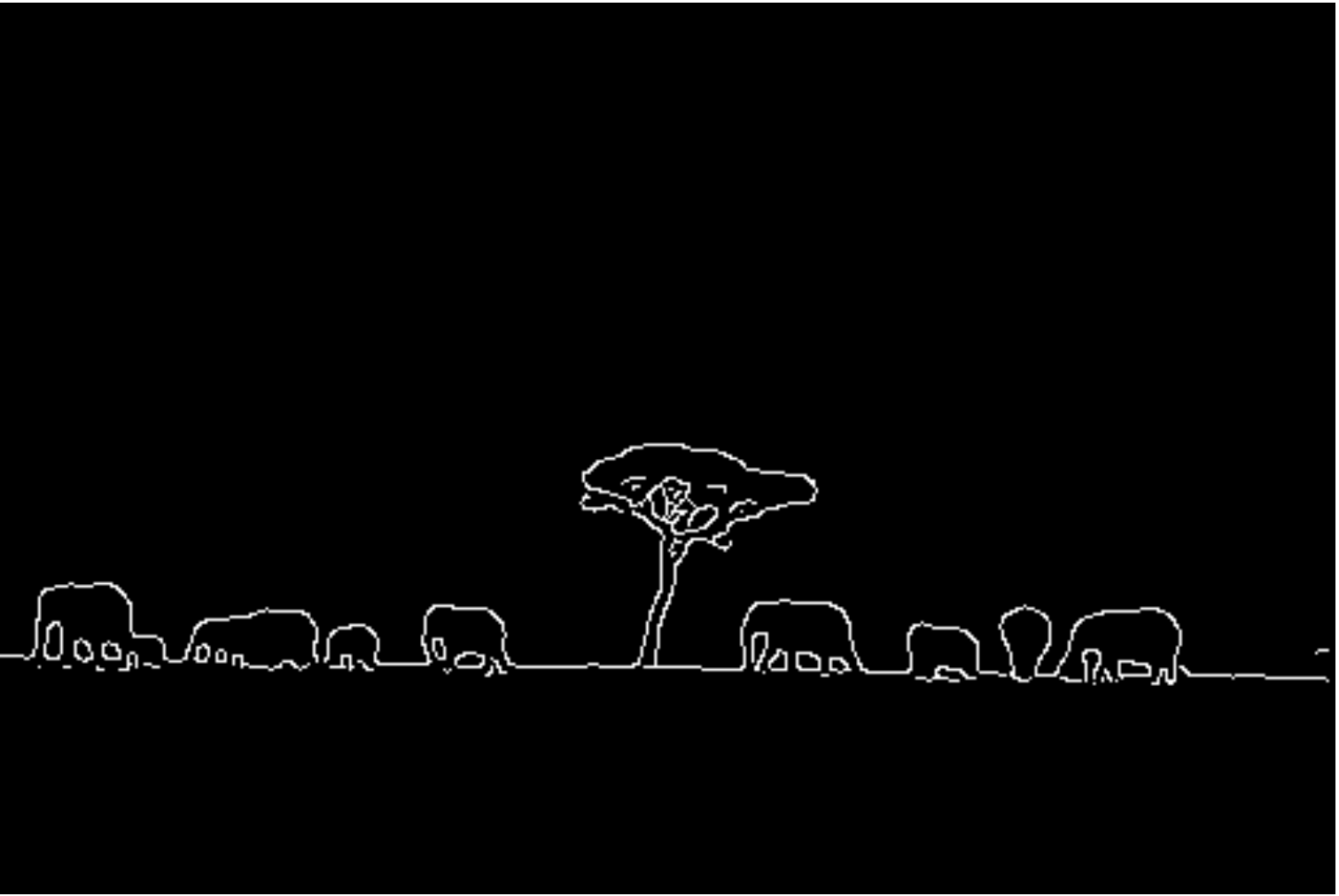}\label{fig:eg1_output_canny}}\\
  \caption{Applying Canny edge detector to the grayscale version of \Cref{edges1}.}
  \label{edges_canny}
\end{figure}

Object textures are sometimes misclassified as edges during the edge detection process. In order to reduce misclassifications, image abstraction as discussed in the last section can be used to suppress object textures. Given an input image as shown in \Cref{fig:eg1_input}, objects of less importance such as clouds and grasses can be eliminated by image abstraction. Using our method, a smooth image as in \Cref{fig:eg1_output} is obtained. Edge detection or segmentation can then be applied to the output image to obtain a result with much fewer distortions. \Cref{edges_canny} shows the results of applying the Canny edge detector to the grayscale version of \Cref{edges1}. We see that while the output of other models keep unnecessary details, our model produces result containing only salient edges, and removes unimportant details.

%%%%%%%%%%%%%%%%%%%%%%%%%%%%%%%%%%%%%%%%%%%%%%%%%%%%%%%%%%%%%%%%%%%%%%%%%%%%%%%%
\subsection{Details exaggeration}
\label{Enhance_histogram}
\begin{figure}[htb]
  \centering
  \subfloat[Input $I$]{\includegraphics[width=0.23\textwidth,clip]{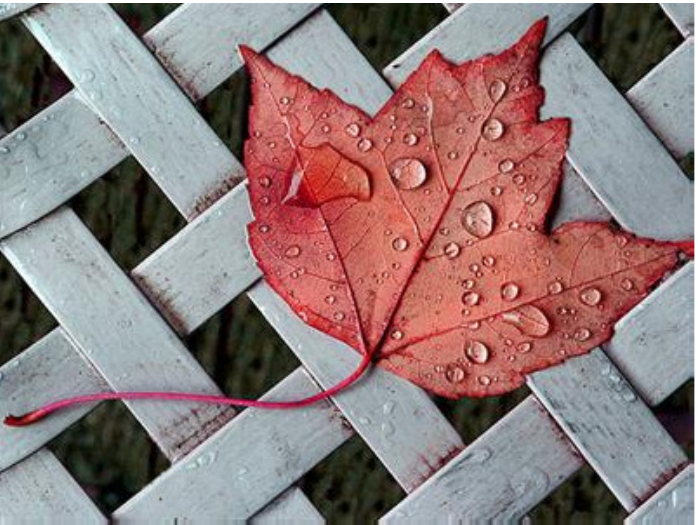}\label{fig:ex1_input}}\,
  \subfloat[][\centering Output $X$ from {\eqref{Proposed_algo}}

  $\lambda = 25$, $\sigma = 0.4$]{\includegraphics[width=0.23\textwidth,clip]{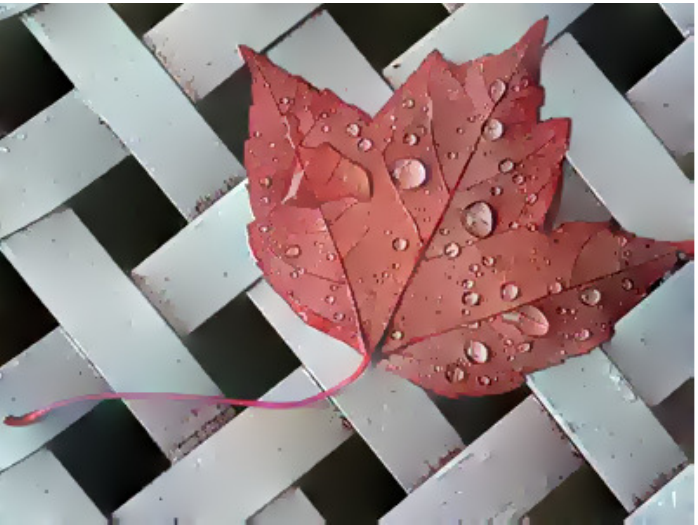}\label{fig:ex1_output}}\,
  \subfloat[][\centering Magnitude of 

  \centering$|I-X|$]{\includegraphics[width=0.23\textwidth,clip]{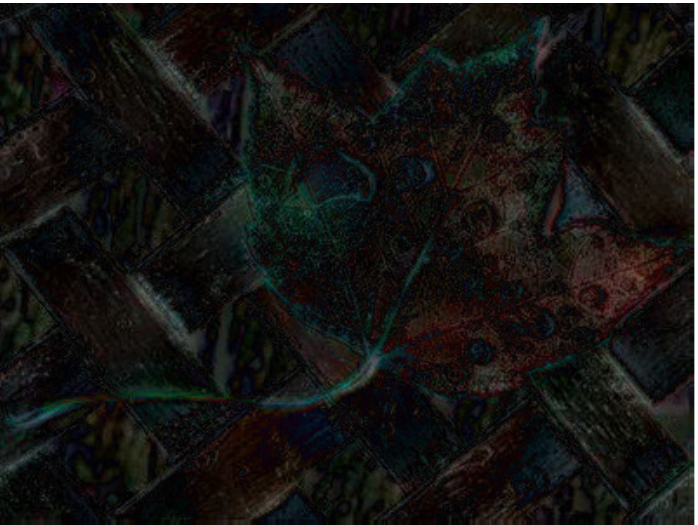}\label{fig:ex1_residue}}\,
  \subfloat[][Details exaggerated

  \centering$J = X + 2(I-X)$]{\includegraphics[width=0.23\textwidth,clip]{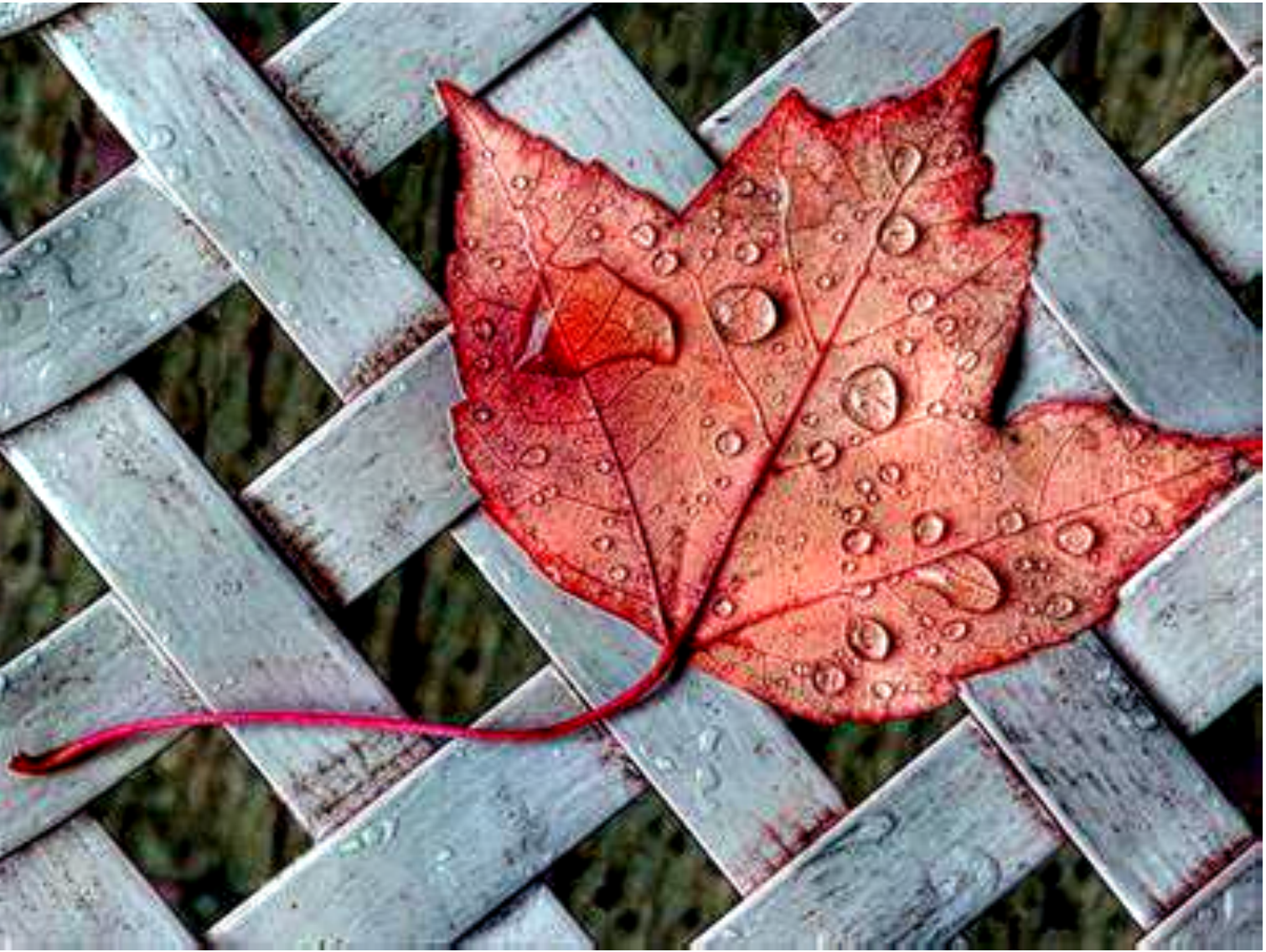}\label{fig:ex1_enhanced}}\\
  \caption{Steps to obtain a details-exaggerated image $J$.}
  \label{exaggeration1}
\end{figure}

\begin{figure}[htb]
  \centering
  \subfloat[Input]{\includegraphics[width=0.3\textwidth,clip]{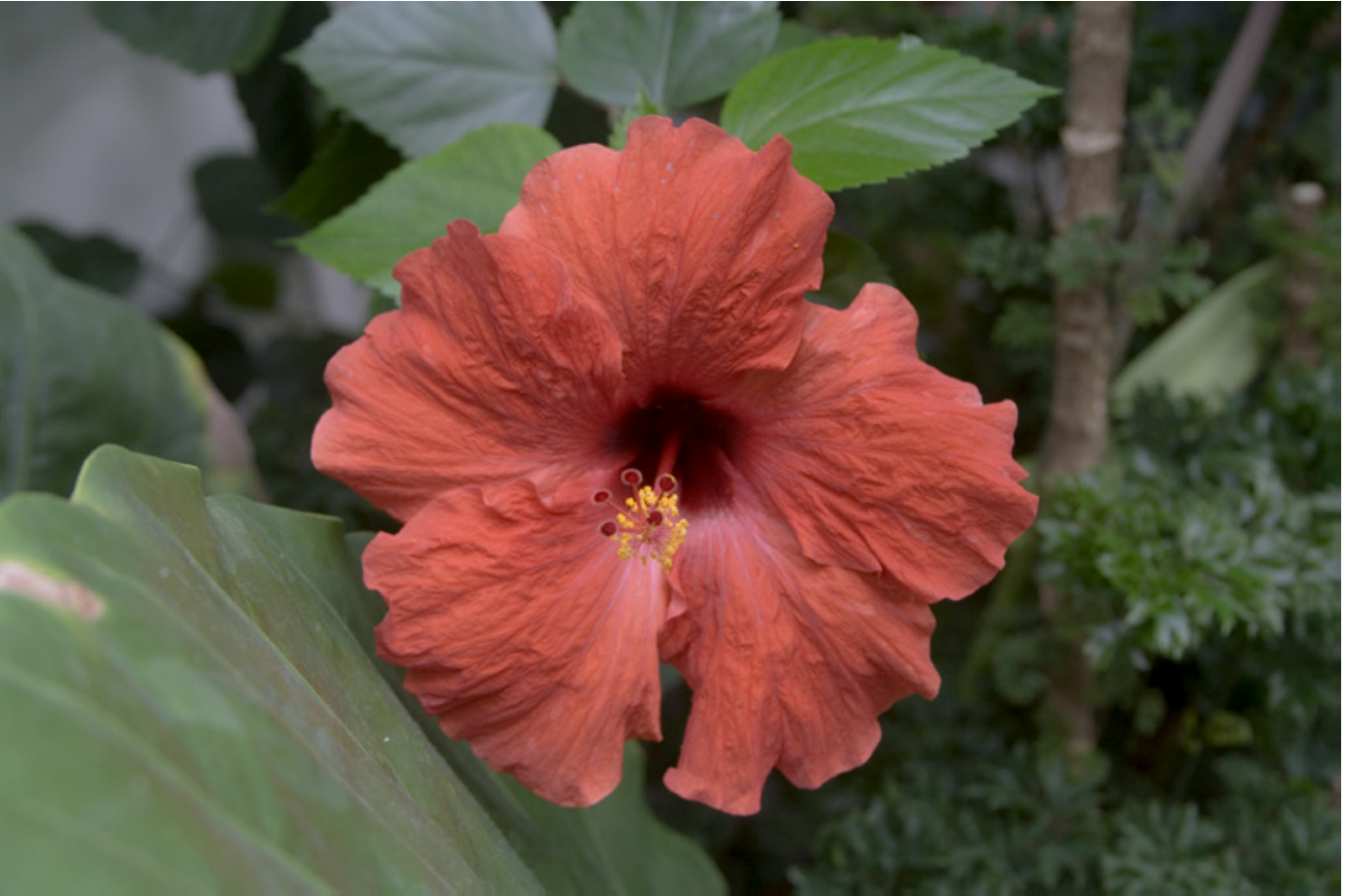}\label{fig:ex2_input}}\,
  \subfloat[][\centering Bilateral \cite{paris2006fast}

  $\sigma_s = 17$, $\sigma_r = 20$, $s = 4$]{\includegraphics[width=0.3\textwidth,clip]{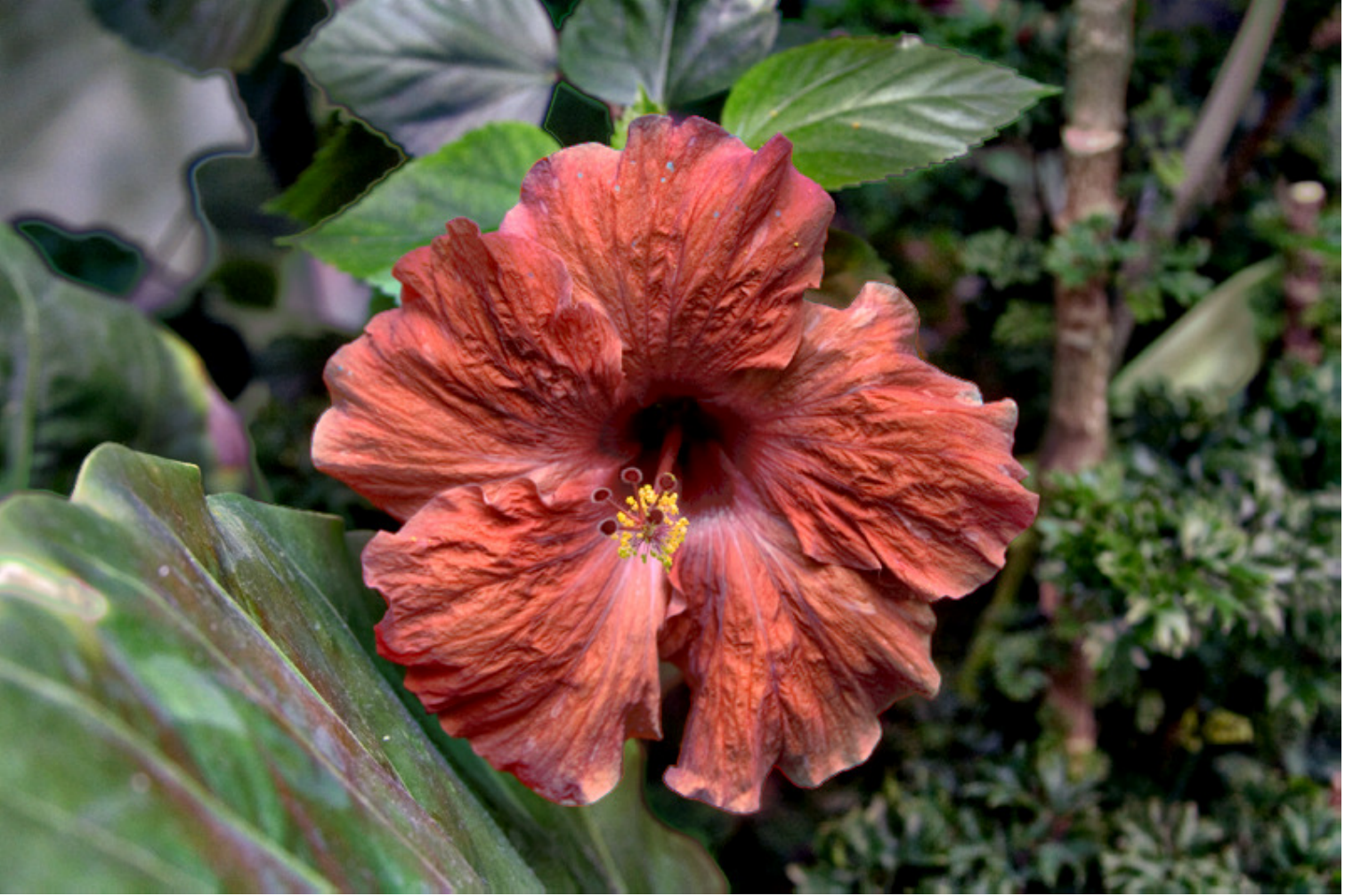}\label{fig:ex2_bilateral}}\,
  \subfloat[WLS \cite{farbman2008edge}]{\includegraphics[width=0.3\textwidth,clip]{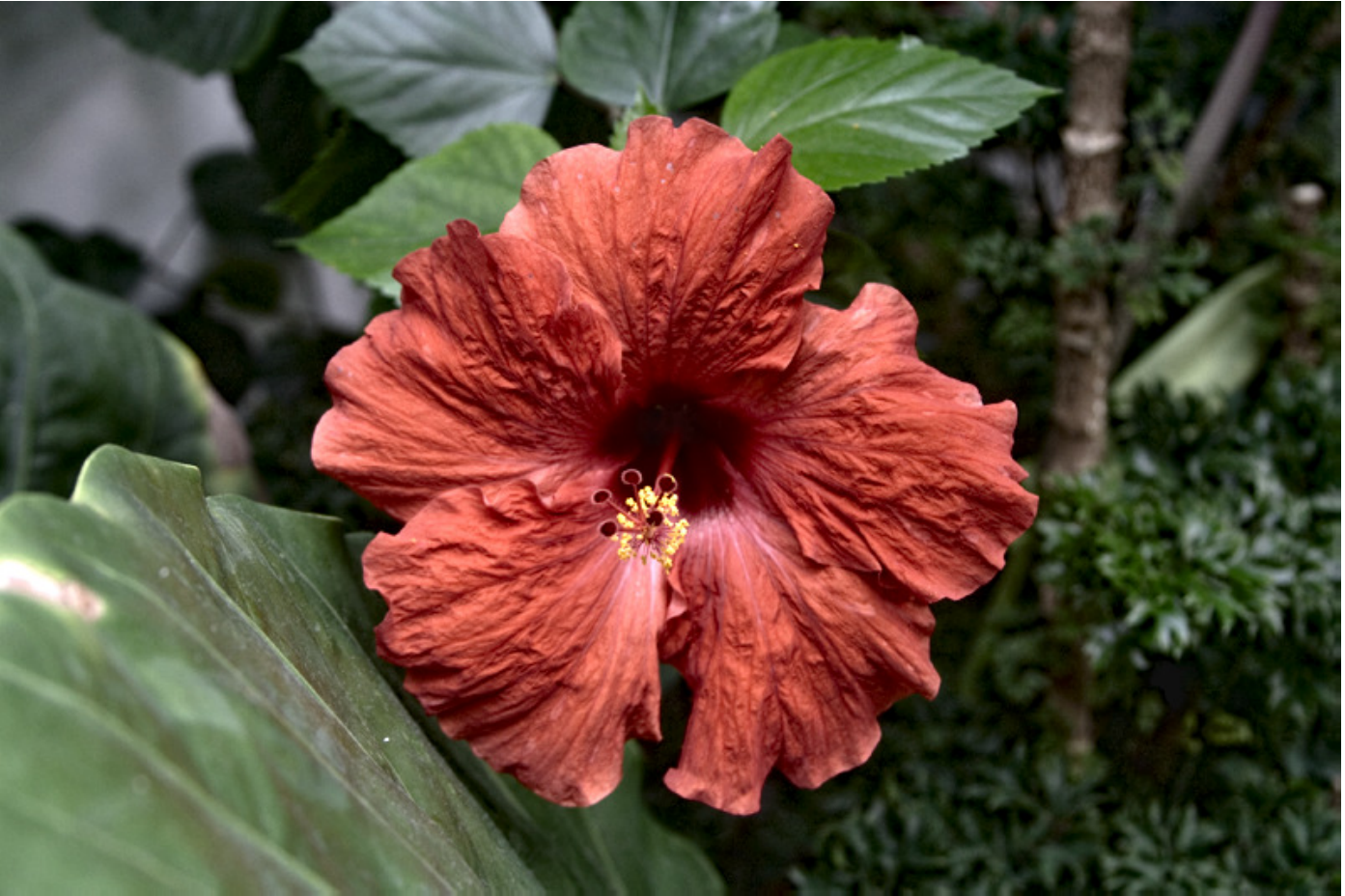}\label{fig:ex2_WLS}}\\
  \subfloat[$l_0$-smoothing \cite{xu2011image}]{\includegraphics[width=0.3\textwidth,clip]{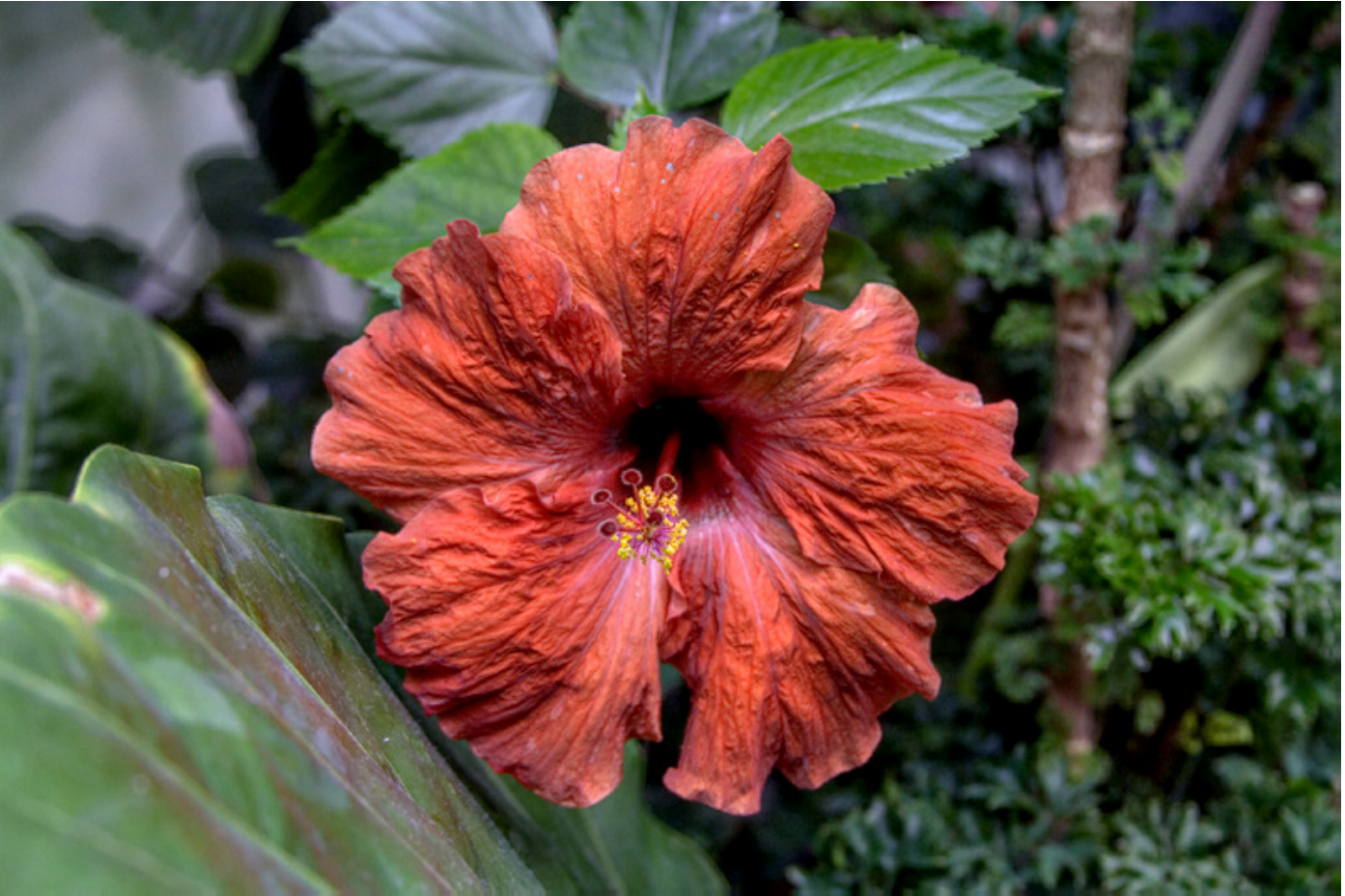}\label{fig:ex2_jia}}\,
  \subfloat[][\centering $l_0$-projection \cite{ono2017l_}

  $\alpha = 127920$, $s = 2$]{\includegraphics[width=0.3\textwidth,clip]{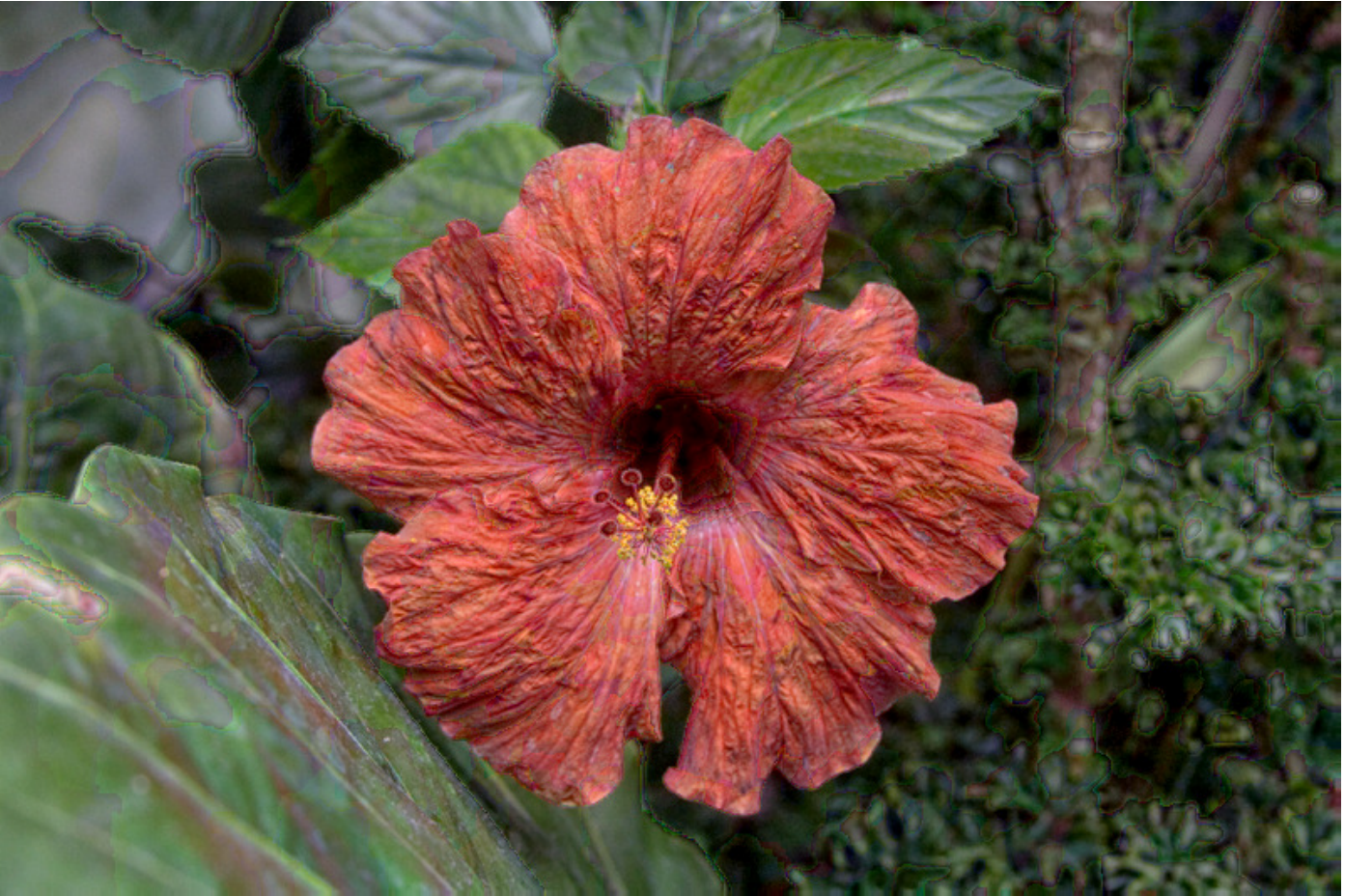}\label{fig:ex2_l0proj}}\,
  \subfloat[][\centering Ours

  \centering$\lambda=13, s = 2.5$, $\sigma = 0$]{\includegraphics[width=0.3\textwidth,clip]{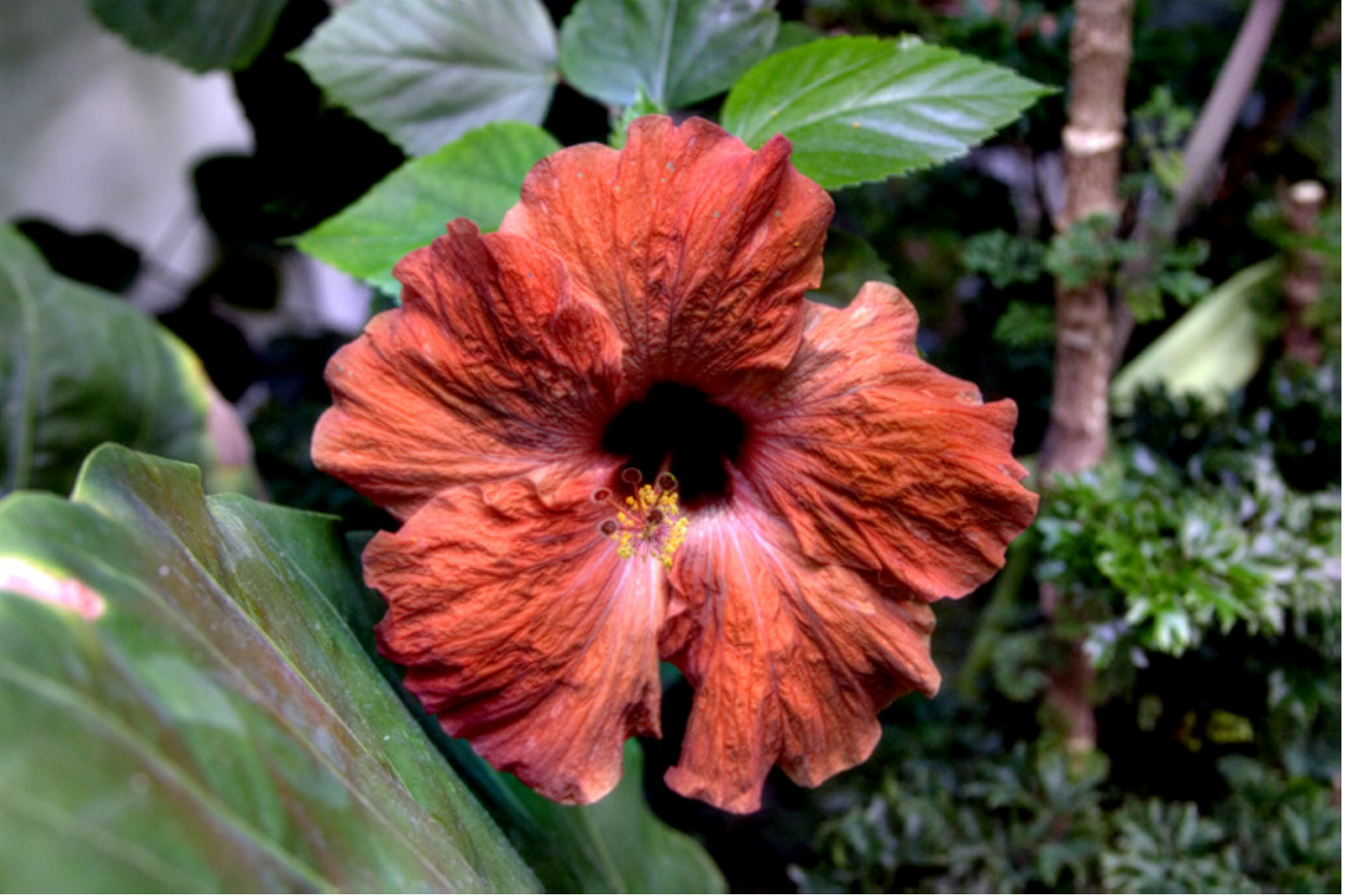}\label{fig:ex2_output}}
  \caption{Comparison of our method with other methods in details exaggeration.}
  \label{exa_com}
\end{figure}

Details exaggeration is to enhance the fine details in an image as much as possible. Given an input image $I$, we obtain a smooth image $X$ by our method where the textures in $I$ are removed, see in \Cref{fig:ex1_output}.  As seen in \Cref{fig:ex1_residue},
the image $|I-X|$ has small values in regions with insignificant textures and large values in the parts containing strong textures. By enhancing ($I-X$) and adding it back to $X$, a details-exaggerated image $J$ can be obtained, see \Cref{fig:ex1_enhanced}. Mathematically, we have $J = X + s(I-X)$, where $s>1$ is a parameter controlling the extent of exaggeration. \Cref{exa_com} shows a comparison with the results by other methods. In \Cref{fig:ex2_output} we see that our model successfully produces better result with more exaggerated details.

%%%%%%%%%%%%%%%%%%%%%%%%%%%%%%%%%%%%%%%%%%%%%%%%%%%%%%%%%%%%%%%%%%%%%%%%%%%%%%%%
\subsection{Scan-through removal}
\label{S:scanthrough}
\begin{figure}[htb]
  	\centering
  	\includegraphics[width=0.32\textwidth,clip]{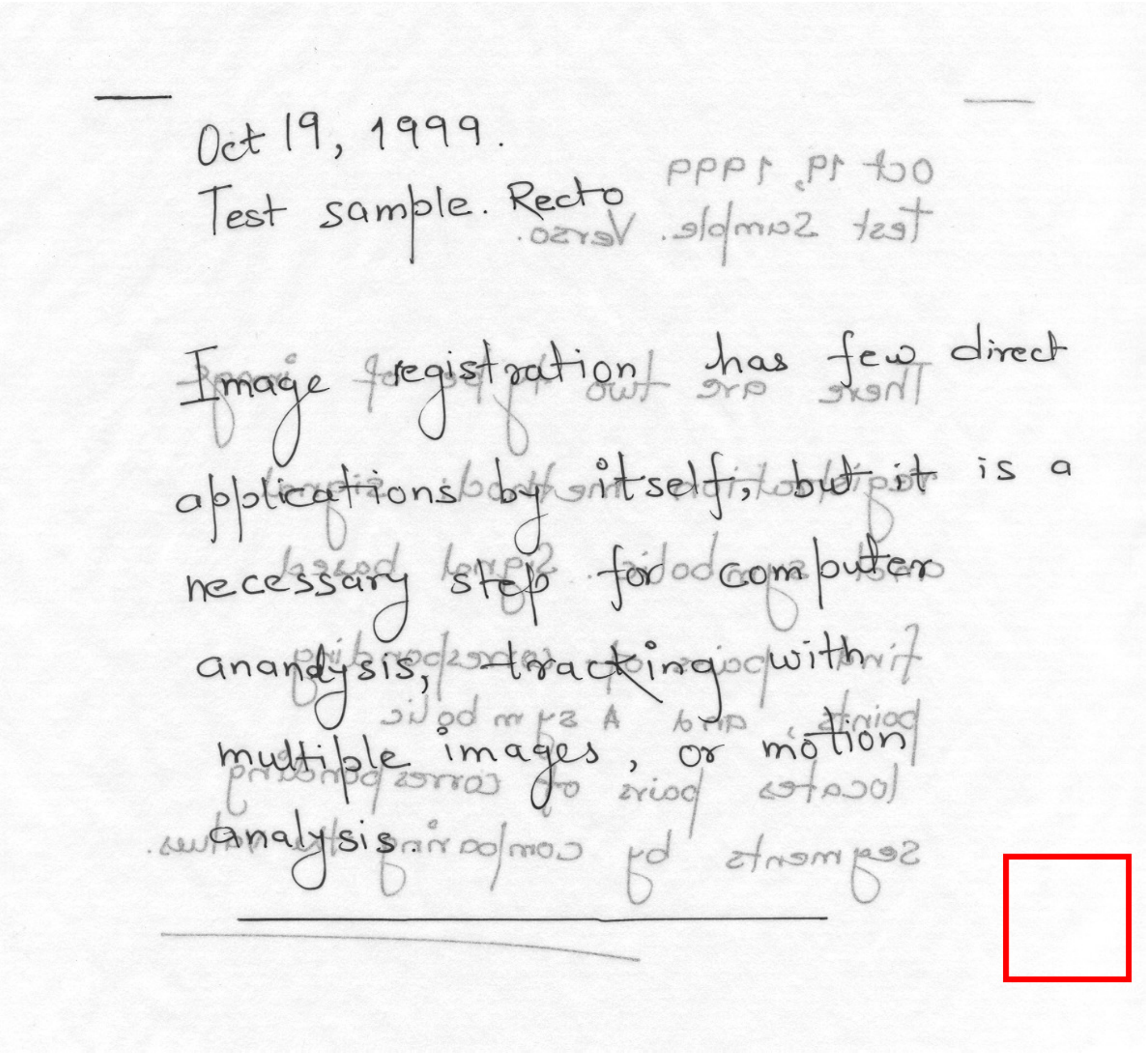}\label{fig:background_detect_1}\,
	\includegraphics[width=0.32\textwidth,clip]{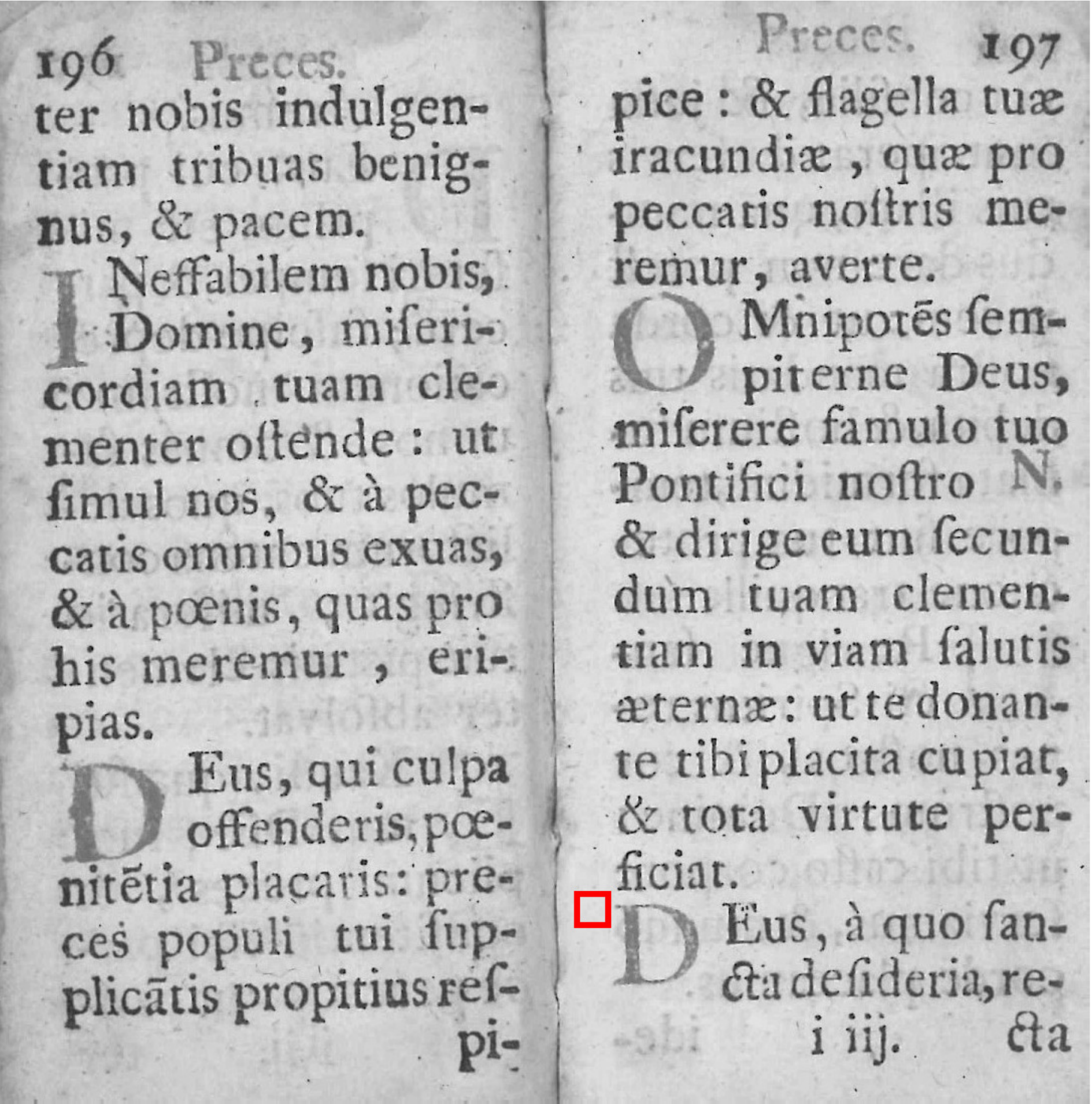}\label{fig:background_detect_2}\, 
	\includegraphics[width=0.32\textwidth,clip]{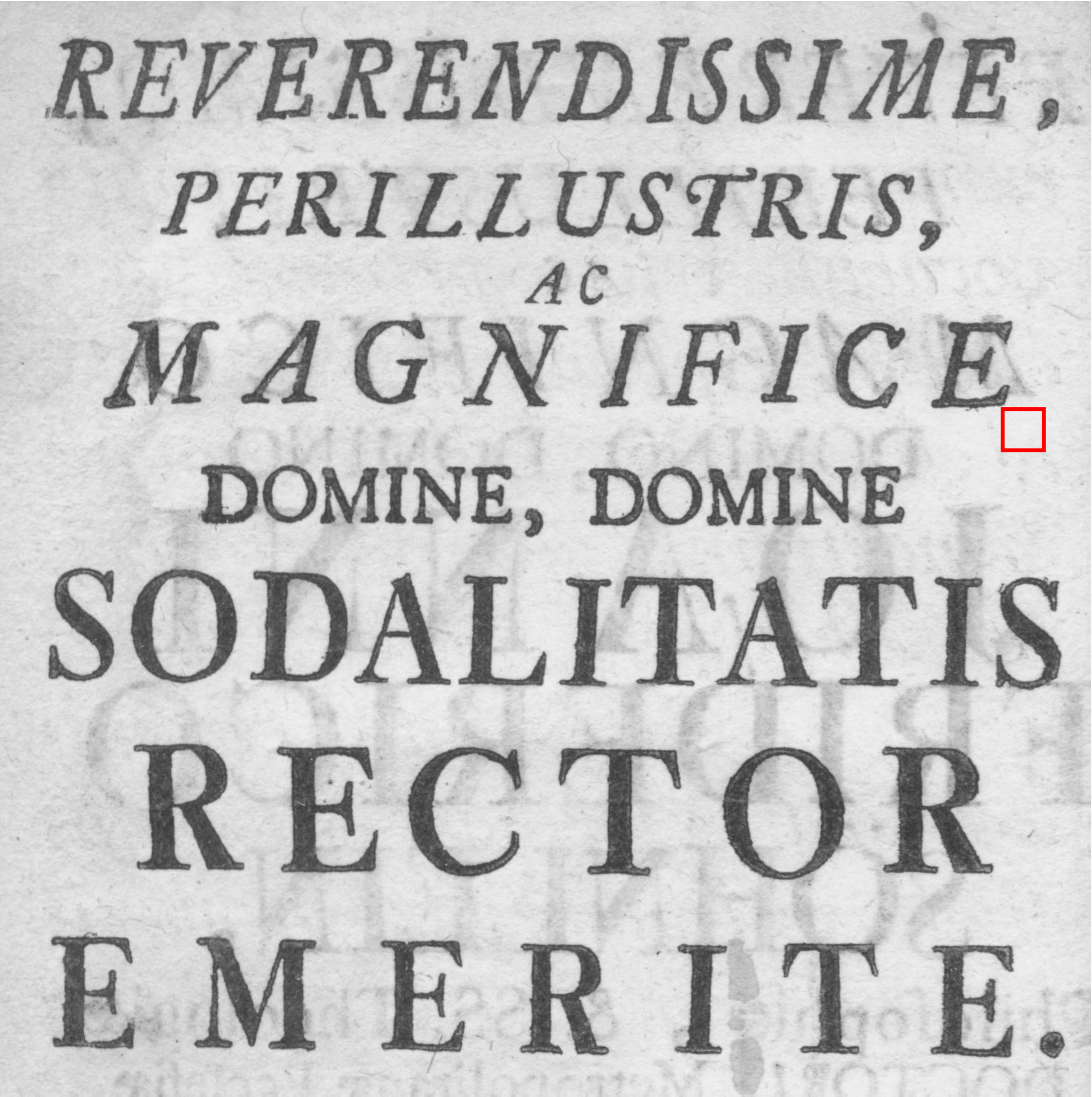}\label{fig:background_detect_3}
  	\caption{Background region detection for three different inputs. Red-colored squares are the regions located by our procedure.}
	\label{background_detection}
\end{figure}

Two-sided documents can be suffered from the effect of back-to-front interference, known as ``see-through". Usually, ``see-through'' produces relatively small gradient fluctuations than the main content we want to preserve, see \Cref{background_detection}. By considering the edges, one can identify interferences and eliminate them.

Recall in \eqref{C_scan}, we also impose a constraint that background pixels will not be modified. Here background pixels refer to the pixels with values not less than $\alpha$.
To find a suitable $\alpha$, we first need to locate background regions---regions which contain only insignificant intensity change, i.e. the standard deviation of the intensity within the region should be small. Motivated by this, we design a multi-scale sliding window method to compute a suitable $\alpha$.  A sliding window with size $w$ is used to scan through an input image $Y$ with stride $\lceil w/5 \rceil$. At each location $p$, the mean intensity $m_p$ of the sliding window is computed and if its standard deviation $\sigma_p$ is smaller than a parameter $\hat{\sigma}$, $m_p$ will be stored for future selection. After scanning through the whole image, we set $\alpha$ to the largest stored value to avoid choosing regions with purely foreground or interference. If $\sigma_p \geq \hat{\sigma}$ for all $p$, we replace $w$ by $w/2$ and scan through the image again. At the worst case when $w=1$, it is equivalent to setting $\alpha$ to the maximum intensity of the image. In our tests, we use $\hat{\sigma} = 3$.

The reason for using a varying window size is that a small region will have a chance of capturing extreme value and a large window will have a chance of failing in capturing purely background. Therefore, we start from {a} large window and stop once we find at least one region with small standard deviation. The initial window we use is the largest square window of length $w = 2^\ell$ that can fit inside the given image.
\Cref{background_detection} shows the windows (red-colored squares) obtained by the procedure above. We see that it successfully locates regions with pure background. The background level $\alpha$ is the mean intensity of the corresponding square.

\begin{figure}[htb]
  \centering
  \subfloat[][\centering Nishida and Suzuki \cite{nishida2003correcting}, 
  $S = 2^7$, $\lambda = 130$]{\includegraphics[width=0.48\textwidth,clip]{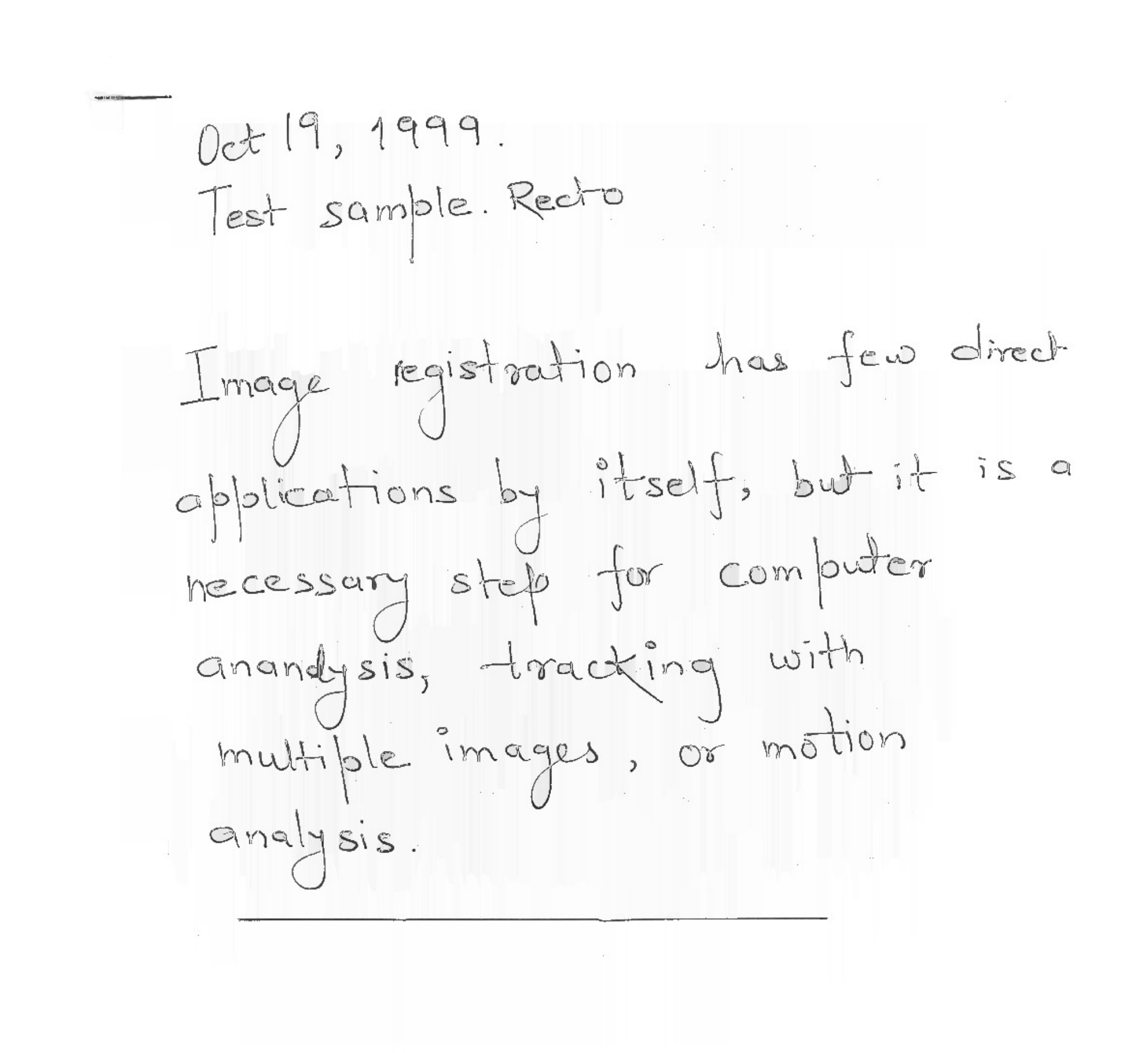}
  \label{fig:scan_Nishida}}\,
  \subfloat[][\centering Ours, 
$\lambda=70, \alpha = 255$, $\sigma = 0$]{\includegraphics[width=0.48\textwidth,clip]{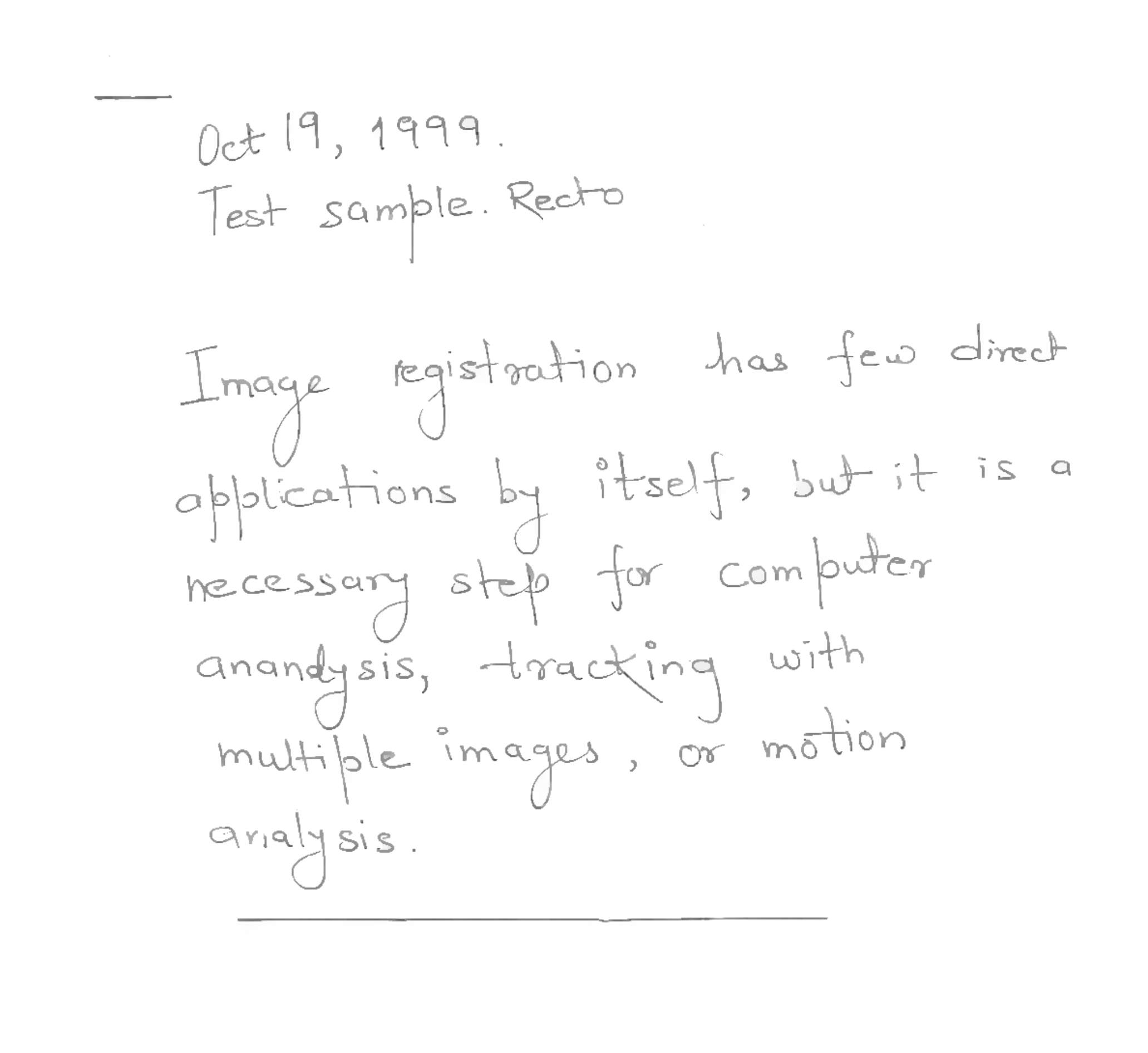}\label{fig:scan_output}}
  \caption{Comparison of our method to a blind method in scan-through removal.}
\label{scan}
\end{figure}

With $\alpha$ found, we solve our model {\eqref{Proposed_algo}} with constraint \eqref{C_scan} to obtain the output. 
We test our method using the first image in \Cref{background_detection}.
Our output is shown in \Cref{fig:scan_output}, where we see that the contents are kept and the back-page interferences are removed. \Cref{fig:scan_Nishida} shows a comparison with the blind method from \cite{nishida2003correcting}. We also compare our result with three non-blind methods \cite{martinelli2012nonlinear, tonazzini2010multichannel, hyvarinen1999fast}. For copyright reasons, we can only refer readers to the papers \cite{tonazzini2010multichannel,martinelli2012nonlinear} to see the resulting images from the three methods. Our method outperforms the blind method and is comparable to the non-blind methods, while these non-blind methods require information from both sides.
%%%%%%%%%%%%%%%%%%%%%%%%%%%%%%%%%%%%%%%%%%%%%%%%%%%%%%%%%%%%%%%%%%%%%%%%%%%%%%%%
%%%%%%%%%%%%%%%%%%%%%%%%%%%%%%%%%%%%%%%%%%%%%%%%%%%%%%%%%%%%%%%%%%%%%%%%%%%%%%%%
\section{Conclusion}
\label{S:5}
We have proposed a convex model with suitable constraints for edge-preserving smoothing tasks including image abstraction, edge extraction, details exaggeration, and documents scan-through removal. Our convex model allows us to solve it efficiently by existing algorithms.

In this paper, because of the special applications we considered, we use only the thresholded histograms as target edge-histograms. In the future, we would investigate more general shapes of edge-histograms and apply them to a wider class of problems.

\bibliographystyle{abbrv}
\bibliography{EdgeHistSpec}

\begin{thebibliography}{10}

\bibitem{beck2009fast}
A.~Beck and M.~Teboulle.
\newblock A fast iterative shrinkage-thresholding algorithm for linear inverse
  problems.
\newblock {\em SIAM Journal on Imaging Sciences}, 2(1):183--202, 2009.

\bibitem{black1998robust}
M.~Black, G.~Sapiro, D.~Marimont, and D.~Heeger.
\newblock Robust anisotropic diffusion.
\newblock {\em IEEE Transactions on Image Processing}, 7(3):421--432, 1998.

\bibitem{chambolle2004algorithm}
A.~Chambolle.
\newblock An algorithm for total variation minimization and applications.
\newblock {\em Journal of Mathematical Imaging and Vision}, 20(1-2):89--97,
  2004.

\bibitem{chen2007real}
J.~Chen, S.~Paris, and F.~Durand.
\newblock Real-time edge-aware image processing with the bilateral grid.
\newblock In {\em ACM Transactions on Graphics}, volume~26, pages 103.1--103.9.
  ACM, 2007.

\bibitem{yu2010otsu}
Y.~Chen, D.~Chen, Y.~Li, and L.~Chen.
\newblock Otsu's thresholding method based on gray level-gradient
  two-dimensional histogram.
\newblock In {\em 2010 2nd International Asia Conference on Informatics in
  Control, Automation and Robotics}, volume~3, pages 282--285. IEEE, 2010.

\bibitem{cheng2014feature}
X.~Cheng, M.~Zeng, and X.~Liu.
\newblock Feature-preserving filtering with l0 gradient minimization.
\newblock {\em Computers \& Graphics}, 38:150--157, 2014.

\bibitem{coltuc2006exact}
D.~Coltuc, P.~Bolon, and J.~Chassery.
\newblock Exact histogram specification.
\newblock {\em IEEE Transactions on Image Processing}, 15(5):1143--1152, 2006.

\bibitem{estrada2009manuscript}
R.~Estrada and C.~Tomasi.
\newblock Manuscript bleed-through removal via hysteresis thresholding.
\newblock In {\em 2009 10th International Conference on Document Analysis and
  Recognition}, pages 753--757. IEEE, 2009.

\bibitem{farbman2008edge}
Z.~Farbman, R.~Fattal, D.~Lischinski, and R.~Szeliski.
\newblock Edge-preserving decompositions for multi-scale tone and detail
  manipulation.
\newblock In {\em ACM Transactions on Graphics}, volume~27, pages 67.1--67.10.
  ACM, 2008.

\bibitem{gabay1976dual}
D.~Gabay and B.~Mercier.
\newblock A dual algorithm for the solution of nonlinear variational problems
  via finite element approximation.
\newblock {\em Computers \& Mathematics with Applications}, 2(1):17--40, 1976.

\bibitem{gerace2016inpainting}
I.~Gerace, C.~Palomba, and A.~Tonazzini.
\newblock An inpainting technique based on regularization to remove
  bleed-through from ancient documents.
\newblock In {\em 2016 International Workshop on Computational Intelligence for
  Multimedia Understanding}, pages 1--5. IEEE, 2016.

\bibitem{glowinski2008lectures}
R.~Glowinski.
\newblock {\em Lectures on Numerical Methods for Non-Linear Variational
  Problems}.
\newblock Springer Science \& Business Media, 2008.

\bibitem{gonzales1992digital}
R.~Gonzales and R.~Woods.
\newblock Digital image processing.
\newblock {\em Addison-Welsley, Reading, MA}, 1992.

\bibitem{hyvarinen1999fast}
A.~Hyvarinen.
\newblock Fast and robust fixed-point algorithms for independent component
  analysis.
\newblock {\em IEEE Transactions on Neural Networks}, 10(3):626--634, 1999.

\bibitem{lee2017color}
S.~Lee and C.~Tseng.
\newblock Color image enhancement using histogram equalization method without
  changing hue and saturation.
\newblock In {\em 2017 IEEE International Conference on Consumer
  Electronics-Taiwan}, pages 305--306. IEEE, 2017.

\bibitem{lim2015new}
S.~Lim, N.~Isa, C.~Ooi, and K.~Toh.
\newblock A new histogram equalization method for digital image enhancement and
  brightness preservation.
\newblock {\em Signal, Image and Video Processing}, 9(3):675--689, 2015.

\bibitem{martinelli2012nonlinear}
F.~Martinelli, E.~Salerno, I.~Gerace, and A.~Tonazzini.
\newblock Nonlinear model and constrained ml for removing back-to-front
  interferences from recto--verso documents.
\newblock {\em Pattern Recognition}, 45(1):596--605, 2012.

\bibitem{merrikh2010using}
F.~Merrikh-Bayat, M.~Babaie-Zadeh, and C.~Jutten.
\newblock Using non-negative matrix factorization for removing show-through.
\newblock In {\em International Conference on Latent Variable Analysis and
  Signal Separation}, pages 482--489. Springer, 2010.

\bibitem{mignotte2012energy}
M.~Mignotte.
\newblock An energy-based model for the image edge-histogram specification
  problem.
\newblock {\em IEEE Transactions on Image Processing}, 21(1):379--386, 2012.

\bibitem{nguyen2015fast}
R.~Nguyen and M.~Brown.
\newblock Fast and effective l0 gradient minimization by region fusion.
\newblock In {\em Proceedings of the IEEE International Conference on Computer
  Vision}, pages 208--216, 2015.

\bibitem{nishida2003correcting}
H.~Nishida and T.~Suzuki.
\newblock Correcting show-through effects on scanned color document images by
  multiscale analysis.
\newblock {\em Pattern recognition}, 36(12):2835--2847, 2003.

\bibitem{ono2017l_}
S.~Ono.
\newblock L0 gradient projection.
\newblock {\em IEEE Transactions on Image Processing}, 26(4):1554--1564, 2017.

\bibitem{pang2015improved}
X.~Pang, S.~Zhang, J.~Gu, L.~Li, B.~Liu, and H.~Wang.
\newblock Improved l0 gradient minimization with l1 fidelity for image
  smoothing.
\newblock {\em PLOS ONE}, 10(9):e0138682, 2015.

\bibitem{paris2006fast}
S.~Paris and F.~Durand.
\newblock A fast approximation of the bilateral filter using a signal
  processing approach.
\newblock In {\em European Conference on Computer Vision}, pages 568--580.
  Springer, 2006.

\bibitem{perona1990scale}
P.~Perona and J.~Malik.
\newblock Scale-space and edge detection using anisotropic diffusion.
\newblock {\em IEEE Transactions on Pattern Analysis and Machine Intelligence},
  12(7):629--639, 1990.

\bibitem{rudin1992nonlinear}
L.~Rudin, S.~Osher, and E.~Fatemi.
\newblock Nonlinear total variation based noise removal algorithms.
\newblock {\em Physica D: Nonlinear Phenomena}, 60(1-4):259--268, 1992.

\bibitem{salerno2013nonlinear}
E.~Salerno, F.~Martinelli, and A.~Tonazzini.
\newblock Nonlinear model identification and see-through cancelation from
  recto--verso data.
\newblock {\em International Journal on Document Analysis and Recognition},
  16(2):177--187, 2013.

\bibitem{savino2016joint}
P.~Savino, L.~Bedini, and A.~Tonazzini.
\newblock Joint non-rigid registration and restoration of recto-verso ancient
  manuscripts.
\newblock In {\em 2016 International Workshop on Computational Intelligence for
  Multimedia Understanding}, pages 1--5. IEEE, 2016.

\bibitem{savino2016digital}
P.~Savino and A.~Tonazzini.
\newblock Digital restoration of ancient color manuscripts from geometrically
  misaligned recto-verso pairs.
\newblock {\em Journal of Cultural Heritage}, 19:511--521, 2016.

\bibitem{sharma2001show}
G.~Sharma.
\newblock Show-through cancellation in scans of duplex printed documents.
\newblock {\em IEEE Transactions on Image Processing}, 10(5):736--754, 2001.

\bibitem{storath2014jump}
M.~Storath, A.~Weinmann, and L.~Demaret.
\newblock Jump-sparse and sparse recovery using potts functionals.
\newblock {\em IEEE Transactions on Signal Processing}, 62(14):3654--3666,
  2014.

\bibitem{sun2016blind}
B.~Sun, S.~Li, X.~Zhang, and J.~Sun.
\newblock Blind bleed-through removal for scanned historical document image
  with conditional random fields.
\newblock {\em IEEE Transactions on Image Processing}, 25(12):5702--5712, 2016.

\bibitem{thomas2008image}
G.~Thomas.
\newblock Image segmentation using histogram specification.
\newblock In {\em 2008 15th IEEE International Conference on Image Processing},
  pages 589--592. IEEE, 2008.

\bibitem{tobias2002image}
O.~Tobias and R.~Seara.
\newblock Image segmentation by histogram thresholding using fuzzy sets.
\newblock {\em IEEE Transactions on Image Processing}, 11(12):1457--1465, 2002.

\bibitem{tomasi1998bilateral}
C.~Tomasi and R.~Manduchi.
\newblock Bilateral filtering for gray and color images.
\newblock In {\em Sixth International Conference on Computer Vision, 1998},
  pages 839--846. IEEE, 1998.

\bibitem{tonazzini2004independent}
A.~Tonazzini, L.~Bedini, and E.~Salerno.
\newblock Independent component analysis for document restoration.
\newblock {\em Document Analysis and Recognition}, 7(1):17--27, 2004.

\bibitem{tonazzini2010multichannel}
A.~Tonazzini, I.~Gerace, and F.~Martinelli.
\newblock Multichannel blind separation and deconvolution of images for
  document analysis.
\newblock {\em IEEE Transactions on Image Processing}, 19(4):912--925, 2010.

\bibitem{tonazzini2007fast}
A.~Tonazzini, E.~Salerno, and L.~Bedini.
\newblock Fast correction of bleed-through distortion in grayscale documents by
  a blind source separation technique.
\newblock {\em International Journal of Document Analysis and Recognition},
  10(1):17--25, 2007.

\bibitem{tonazzini2015non}
A.~Tonazzini, P.~Savino, and E.~Salerno.
\newblock A non-stationary density model to separate overlapped texts in
  degraded documents.
\newblock {\em Signal, Image and Video Processing}, 9(1):155--164, 2015.

\bibitem{wang1999image}
Y.~Wang, Q.~Chen, and B.~Zhang.
\newblock Image enhancement based on equal area dualistic sub-image histogram
  equalization method.
\newblock {\em IEEE Transactions on Consumer Electronics}, 45(1):68--75, 1999.

\bibitem{weiss2006fast}
B.~Weiss.
\newblock Fast median and bilateral filtering.
\newblock In {\em ACM Transactions on Graphics}, volume~25, pages 519--526.
  ACM, 2006.

\bibitem{wolf2010document}
C.~Wolf.
\newblock Document ink bleed-through removal with two hidden markov random
  fields and a single observation field.
\newblock {\em IEEE Transactions on Pattern Analysis and Machine Intelligence},
  32(3):431--447, 2010.

\bibitem{xu2011image}
L.~Xu, C.~Lu, Y.~Xu, and J.~Jia.
\newblock Image smoothing via l0 gradient minimization.
\newblock In {\em ACM Transactions on Graphics}, volume~30, pages
  174.1--174.12. ACM, 2011.

\end{thebibliography}
\end{document}